\documentclass[journal,a4paper,final,twocolumn,twoside]{IEEEtran}%
\usepackage{amssymb}
\usepackage{amsmath}
\usepackage{amsfonts}
\usepackage{graphicx}
\usepackage{theorem}
\newtheorem{theorem}{Theorem}

\newtheorem{corollary}{Corollary}

\newtheorem{lemma}{Lemma}

\newtheorem{proposition}{Proposition}

\newtheorem{remark}{Remark}

\begin{document}

\title{Parallelized Structures for MIMO\ FBMC\ under Strong Channel Frequency Selectivity}
\author{Xavier Mestre\thanks{* Corresponding author.}, David Gregoratti\thanks{
Copyright \copyright 2015 IEEE. Personal use of this material is permitted. However, permission to use this material for any other purposes must be obtained from the IEEE by sending a request to pubs-permissions@ieee.org.
This work was partially supported by the European Commission under proyect ICT-318362 EMPhAtiC, and
the Spanish and Catalan Governments under grants TEC2014-59255-C3-1 and
2014-SGR-1567. The material in this paper has been partially presented at
IEEE\ ICASSP 2014. The authors are with the Centre Tecnol\`{o}gic de Telecomunicacions de Catalunya, Av. de Carl Friedrich
Gauss, 7, 08860 Castelldefels (Barcelona), Spain, Phone: +34 936452900, Fax: +34 936452901, E-mail: \{xavier.mestre,david.gregoratti\}@cttc.cat}}
\maketitle

\begin{abstract}%

A novel architecture for MIMO transmission and reception of filterbank
multicarrier (FBMC) modulated signals under strong frequency selectivity is
presented. The proposed system seeks to approximate an ideal
frequency-selective precoder and linear receiver by Taylor expansion,
exploiting the structure of the analysis and synthesis filterbanks. The
resulting architecture is implemented by linearly combining conventional MIMO
linear transceivers, which are applied to sequential derivatives of the
original filterbank. The classical per-subcarrier precoding/linear receiver
configuration is obtained as a special case of this architecture, when only
one stage is fixed at both transmitter and receiver. An asymptotic expression
for the resulting intersymbol/intercarrier (ISI/ICI) distortion is derived
assuming that the number of subcarriers grows large. This expression can in
practice be used in order to determine the number of parallel stages that need
to be implemented in the proposed architecture. Performance evaluation studies
confirm the substantial advantage of the proposed scheme in practical
frequency-selective MIMO scenarios.%
\end{abstract}


\section{Introduction}

The increasing demand for high data rate wireless services has recently
motivated a renewed interest in spectrally efficient signalling methodologies
in order to overcome the current spectrum scarcity. In this context,
filterbank multicarrier (FBMC) modulations have become very strong candidates
to guarantee an optimum spectrum usage while maintaining the nice processing
properties of multicarrier signals, such as reduced complexity equalization.
Unlike cyclic-prefix OFDM\ (CP-OFDM), FBMC\ modulations do not require the use
of a cyclic prefix and can be constructed via spectrally contained pulse
shaping architectures. This significantly increases the spectral efficiency of
the system, improves the spectral localization of the transmitted signal and
reduces the need for guard bands. FBMC\ modulations can be combined with
multi-antenna MIMO technology in order to boost the link system capacity,
leading to an extremely high spectral efficiency.

Even though several alternative FBMC\ modulation formats have been proposed
over the last few years, the most interesting one from the point of view of
spectral efficiency remains to be FBMC based on offset QAM (FBMC/OQAM)
\cite{Saltzberg67,siohan02}. This modulation is constructed via a critically
sampled uniformly spaced\ filterbanks modulated by real-valued\ symbols. Given
the fact that there is no CP and since the filterbanks are critically sampled,
FBMC/OQAM achieves the largest possible spectral efficiency in the whole class
of FBMC\ modulations. Furthermore, by conveniently selecting the prototype
filters at the transmit and receive side, one can perfectly recover the
transmitted symbols at the receiver in the presence of a noiseless frequency
flat channel. For this reason, this modulation is widely considered as the
most prominent FBMC modulation.

A second important class of FBMC modulations are typically referred to as
Filtered Multi Tone (FMT) or FBMC/QAM \cite{Hleiss97,Cherubini02}. The idea
behind these modulations consists in directly modulating complex\ QAM symbols
instead of real-valued ones, avoiding again the introduction of a CP. The
approach is clearly more versatile than FBMC/OQAM and since the signalling is
carried out on complex symbols, the modulation operative becomes very similar
to classical CP-OFDM. However, it can be seen that if the filterbank is
critically sampled it is not possible to perfectly recover the transmitted
symbols, even in the presence of a noiseless frequency flat channel. For this
reason, FBMC/QAM is typically implemented using oversampled
filterbanks\footnote{By critical sampling we mean that the signals at the
input of the synthesis filterbank is interpolated by a factor that is equal to
the number of subcarriers. When the interpolation factor is higher than the
number of subcarriers, we say that the filterbank is oversampled (also
overinterpolated). FBMC/QAM typically uses oversampling ratios between 5/4 and
3/2 \cite{Siclet2006, Hleiss97}, which may be comparable to the efficiency
loss in CP-OFDM due to the insertion of the cyclic prefix.}, which clearly
reduce the spectral efficiency of the system. As a consequence of using
oversampled filterbanks, the different constituent filters have little overlap
in the frequency domain, which minimizes potential problems in terms of
inter-carrier interference (ICI) with respect to FBMC/OQAM. For all these
reasons, FBMC/QAM is today a valid alternative employed in commercial systems
such as the professional mobile radio system TETRA\ Enhanced Data System
(TEDS) \cite{TEDS_PHY_2011}. However, due to the introduction of redundancy
(oversampling) in the transmit/receive filterbanks, FBMC/QAM can only achieve
a portion of the spectral efficiency of a classical FBMC/OQAM modulation.

Several other alternative FBMC\ modulations have been proposed in the
literature, although they typically rely on the introduction of CP, which
simplifies the equalization but clearly incurs in a significant spectral loss.
This CP is not needed in FBMC modulations, because under relatively mild
channel frequency selectivity the channel response can be assumed to be
approximately flat within each subcarrier band. Hence, a single-tap
per-subcarrier weighting is in principle sufficient to equalize the system, as
it is the case in CP-OFDM. Unfortunately, in the presence of strong channel
frequency selectivity, the channel can no longer be approximated as flat
within each subcarrier pass band, and FBMC modulations require more
sophisticated equalization systems (see e.g. \cite{ihalainen07} and references
therein for a review of FBMC\ equalization techniques). In practical terms, if
the receiver keeps using a single-tap per-subcarrier equalizer in the presence
of a highly frequency-selective channel, its output will appear contaminated
by a residual distortion superposed to the background noise. In FBMC/QAM
modulations, this distortion will eminently be related to the
inter-symbol-interference (ISI) caused by the channel within each subband. In
FBMC/OQAM modulations, the strong overlap between the different subband
filters will result in both ICI\ and ISI at the output in the presence of a
frequency-selective channel. Consequently, the effect of channel frequency
selectivity becomes more problematic in FBMC/OQAM modulations, a fact that has
prevented a more widespread acceptance of this modulation in spite of its
clear superiority in terms of spectral efficiency.

This residual distortion under channel frequency selectivity is much more
devastating in MIMO transmissions, basically due to a superposition effect of
the multiple parallel antennas/streams \cite{Estella2010, phydyasD4.2,
Payaro2010}. This incremental distortion effect in MIMO contexts has
traditionally been mitigated using complex receiver strategies, such as
sophisticated equalization architectures \cite{Kofidis2010, Ihalainen2011,
PhydyasD4.1}, or algorithms based on successive interference cancellation
\cite{Ikhlef2009, Tabach2007, PhydyasD4.1}. More recent approaches have
additionally considered the optimization of the transmitter architecture in
order to mitigate the effect of the channel frequency selectivity. For
example, \cite{Caus2012} considers the optimization of the precoder/linear
receiver pair in order to achieve spatial diversity while minimizing the
residual distortion at the output of the receiver. A related approach can be
found in \cite{Moret2011, Weiss2011}, where a polynomial-based (multi-tap) SVD
precoder is applied together with an equivalent multi-tap equalizer at the receiver.

Here we take an approach similar to the one in \cite{mestre13tsp} and propose
a general architecture that can be used to implement multiple
MIMO\ transceivers (precoder plus linear receiver) in highly
frequency-selective channels. Our approach is substantially different from the
one in \cite{Caus2012, Moret2011, Weiss2011}, because rather than focusing on
a particular objective to optimize the transceiver, the proposed architecture
provides a general framework that can be used to construct a variety of MIMO
transceivers. On the other hand, the results in this paper generalize
\cite{mestre13tsp} in several important aspects, even under the
SISO\ configuration. First, the approach in \cite{mestre13tsp} considers the
specific case where the transmitter does not apply any
precoding/pre-equalization processing, while the receiver performs direct
channel inversion. Here, we consider a much more general setting with a
generic precoder/pre-equalizer at the transmitter together with the
corresponding equalizer at the receiver. Second, the asymptotic performance
analysis in \cite{mestre13tsp} is based on the assumption that the prototype
pulses are perfect reconstruction (PR) filters. Here, the analysis is
generalized to the case where the prototypes are not necessarily PR, which is
typically the case in practice. Finally, the analysis in \cite{mestre13tsp} is
only valid for finite impulse response (FIR) channel models. Here, the
analysis is generalized to more general channel forms, not necessarily having
finite impulse response.

Before going into the technical development, it is worth pointing out that the
present study assumes linear, time invariant and perfectly estimated channel
responses. These assumptions are not perfectly met in a practical situation,
mainly because of the presence of amplifier nonlinearities, Doppler effects
and the use of finite training sequences. Still, we assume that all these
imperfections are negligible for the sake of analytical tractability. The
detrimental effect of these nonidealities could in principle be reduced by
increasing the cost of the power amplifier and by employing channel estimates
are refreshed frequently enough and obtained with a sufficiently large
training sequence. However, in practice these ideal conditions do not hold,
and therefore some degradation in the performance should be expected. Previous
analyses establish that the negative effect of these non-idealities in FBMC is
similar to the one in CP-OFDM \cite{Bouhadda14, Roque12, Kofidis13}, which
leads us to believe that the associated performance degradation will not be dramatic.

The rest of the paper is organized as follows. Section \ref{sec:sigmod}
presents the general MIMO\ signal model and the ideal frequency-selective
transceiver that is considered in this paper and Section
\ref{sec:proposedApproach} presents the proposed parallel multi-stage approach
for general FBMC\ systems. The asymptotic performance of the proposed MIMO
architecture is analyzed in Section \ref{sec:performance_analysis} for general
FBMC/OQAM modulations under the assumption that the number of subcarriers
grows large. Finally, Section \ref{sec:numerical} provides a numerical
evaluation of the multi-stage technique and Section \ref{sec:conclusions}%
\ concludes the paper. All technical derivations have been relegated to the appendices.

\section{\label{sec:sigmod}Signal model}

We consider a MIMO system with $N_{T}$ transmit and $N_{R}$ receive antennas.
Let $\mathbf{H}(\omega)$ denote an $N_{R}\times N_{T}$ matrix containing the
frequency response of the MIMO channels, so that the $\left(  i,j\right)  $th
entry of $\mathbf{H}(\omega)$ contains the frequency response between the
$j$th transmit and the $i$th receive antennas. We assume that the MIMO system
is used for the transmission of $N_{S}$ parallel signal streams, $1\leq
N_{S}\leq\min\left\{  N_{R},N_{T}\right\}  $, which correspond to
FBMC\ modulated signals. More specifically, we will denote by $\mathbf{s}%
(\omega)$ an $N_{S}\times1$ column vector that contains the frequency response
of the signal transmitted at each of the $N_{S}$ parallel streams. Hence, each
entry of the vector $\mathbf{s}(\omega)$ is the Fourier transform of a FBMC
modulated symbol stream.

\begin{figure*}[t]
\begin{center}
\includegraphics[width=10cm]{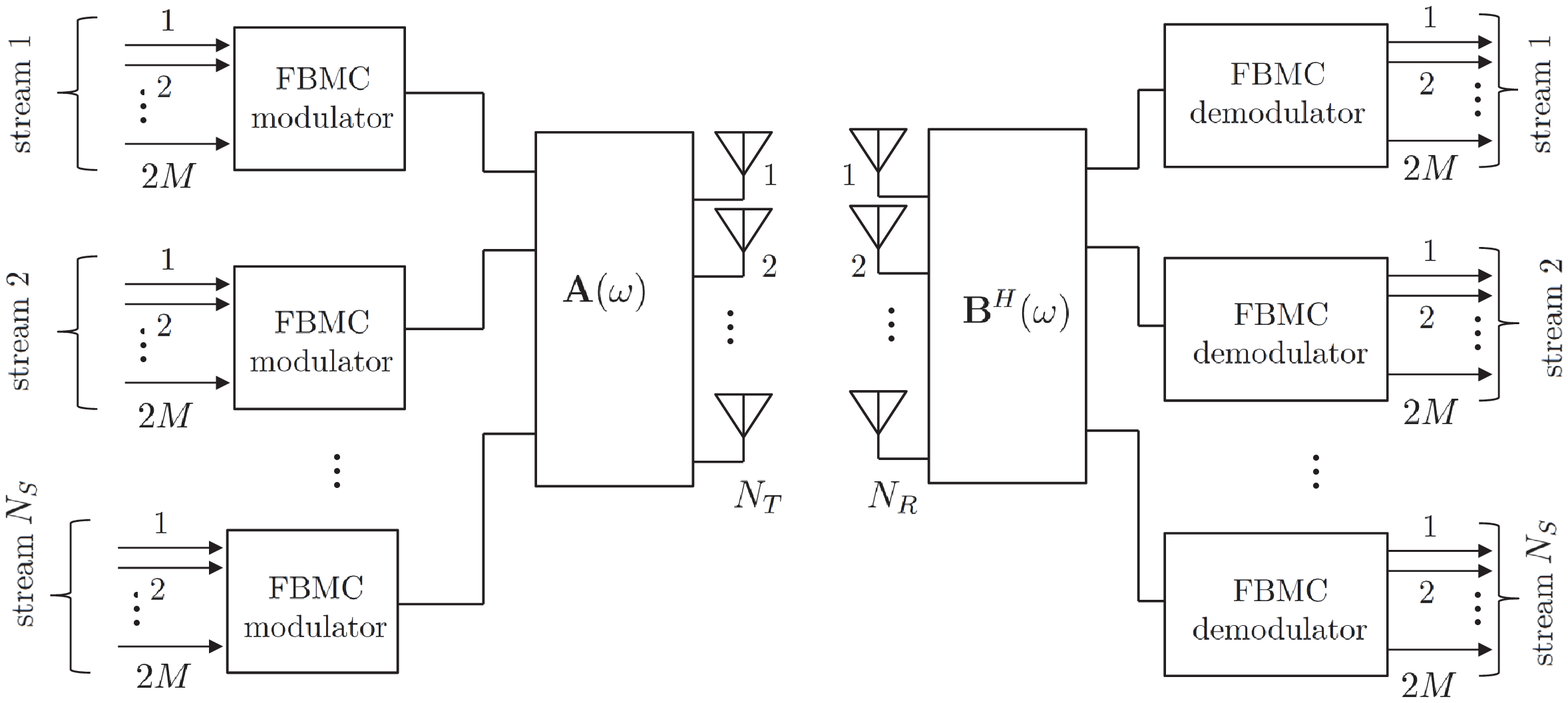}
\end{center}
\caption{Ideal implementation of a frequency selective precoder $\mathbf
{A}(\omega)$ and a linear receiver $\mathbf{B}(\omega
)$ in a FBMC modulation system with $2M$ subcarriers.}
\label{fig:ideal_precoder_receiver}
\end{figure*}

Let us assume that the transmitter applies a frequency-dependent linear
precoder, which will be denoted by the $N_{T}\times N_{S}$ matrix
$\mathbf{A}(\omega)$. The signal transmitted through the $N_{T}$ transmit
antennas can be expressed as
\begin{equation}
\mathbf{x}(\omega)=\mathbf{A}(\omega)\mathbf{s}(\omega) \label{eq:x_omega}%
\end{equation}
where $\mathbf{x}(\omega)$ is an $N_{T}\times1$ column vector containing the
frequency response of the transmitted signal. On the other hand, let
$\mathbf{y}(\omega)$ denote an $N_{R}\times1$ column vector containing the
frequency response of the received signals in noise, namely
\[
\mathbf{y}(\omega)=\mathbf{H}(\omega)\mathbf{x}(\omega)+\mathbf{n}(\omega)
\]
where $\mathbf{n}(\omega)$ is the additive Gaussian white noise. We assume
that the receiver estimates the transmitted symbols by linearly transforming
the received signal vector $\mathbf{y}(\omega)$. More specifically, we
consider a certain $N_{R}\times N_{S}$ receive matrix $\mathbf{B}(\omega)$ so
that the symbols are estimated by
\[
\mathbf{\hat{s}}(\omega)=\mathbf{B}^{H}(\omega)\mathbf{y}(\omega).
\]
The whole ideal frequency-selective transceiver chain is implemented in Fig.
\ref{fig:ideal_precoder_receiver} for a FBMC-modulated system. The number of
subcarriers is fixed to be even, and will be denoted by $2M$.

\begin{figure*}[thb]
\begin{center}
\includegraphics[width=10cm]{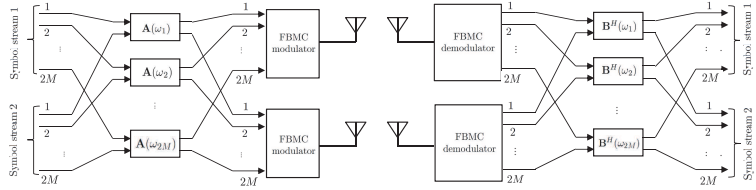}
\end{center}
\caption
{Traditional implementation of the frequecy-selective linear MIMO transmitter and receiver in multicarrier modulations, for the specific case of $N_{T}%
=2$ transmit antennas, $N_{R}=2$ receive antennas and $2M$ subcarriers.}
\label{fig:trad_MIMO_TXRX}
\end{figure*}

The main problem with the MIMO architecture presented in Fig.
\ref{fig:ideal_precoder_receiver} comes from the fact that, in practice, the
frequency-dependent matrices $\mathbf{A}(\omega),$\ $\mathbf{B}(\omega)$ need
to be implemented using real filters. However, these filters have very large
(or even infinite) impulse responses, which may be difficult to implement in
practice. This can partly be solved in multicarrier modulations, as long as it
can be assumed that the frequency selectivity is not severe, so that the
channel response is approximately flat at on each subcarrier pass band. When
this is the case, one can construct the MIMO precoder/receiver operations by
applying the matrices $\mathbf{A}(\omega_{k}),$\ $\mathbf{B}(\omega_{k})$ to
each subcarrier stream, where here $\omega_{k}$ denotes the central frequency
associated with the $k$th subcarrier. This is further illustrated in Fig.
\ref{fig:trad_MIMO_TXRX} for the particular case of $N_{T}=N_{R}=2$ antennas
in a FBMC modulation transmission. Observe that the traditional
(per-subcarrier) implementation in Fig. \ref{fig:trad_MIMO_TXRX} is the result
of changing the position of the precoder/linear receiver with respect to the
FBMC modulator/demodulator in the ideal implementation.

As pointed out above, the traditional solution essentially relies on the fact
that the channel frequency selectivity is mild enough to guarantee that each
sub-carrier observes a frequency non-selective channel. However, under severe
frequency selectivity, the system will suffer from a non-negligible distortion
that will critically impair the performance of the MIMO system. This will be
confirmed below, both analytically and via simulations. Next, we propose an
alternative solution that tries to overcome this effect.

\section{\label{sec:proposedApproach}Proposed approach}

In this section we propose an alternative solution that, with some additional
complexity, significantly mitigates the distortion caused by the channel
frequency selectivity. We assume that the transmit and receive filterbanks are
constructed by modulating a given prototype filter, which may be different at
the transmit and receive sides. We will denote as $p[n]$ and $q[n]$ the
real-valued impulse responses of the transmit and receive prototype pulses,
with Fourier transforms respectively denoted by $P\left(  \omega\right)  $ and
$Q\left(  \omega\right)  $, $\omega\in\mathbb{R}/2\pi\mathbb{Z}$. Hence, the
frequency response of the $k$th filter in the transmit filterbank is assumed
to be equal to $P\left(  \omega-\omega_{k}\right)  \,$\ where $\omega
_{1},\ldots,\omega_{2M}$ denote the subcarrier frequencies, assumed to be
equispaced along the transmitted bandwidth, namely $\omega_{k}=2\pi\left(
k-1\right)  /\left(  2M\right)  $. The following approach could also be
applied to the situation where $P\left(  \omega\right)  $ is different at each
subcarrier, as well as in situations where these subcarriers are not
equispaced. However we prefer to concentrate on the simpler case of uniform
filterbanks to simplify the exposition.

Let us consider the combination of the FBMC\ transmission scheme with the
frequency-selective MIMO precoder $\mathbf{A}(\omega)$. More specifically,
consider the $k$th subcarrier associated with the $n_{S}$th MIMO signal stream
that is sent through the $n_{T}$th transmit antenna. Assuming that the
\textit{ideal} frequency-selective precoder matrix $\mathbf{A}(\omega)$ is
implemented (Fig. \ref{fig:ideal_precoder_receiver}), this stream will
effectively go through a transmit linear system with equivalent frequency
response proportional to%
\[
P\left(  \omega-\omega_{k}\right)  \left\{  \mathbf{A}(\omega)\right\}
_{n_{T},n_{S}}.
\]
The traditional (per-subcarrier) implementation of this precoder is based on
the assumption that $\mathbf{A}(\omega)$ is almost flat along the bandwidth of
$P\left(  \omega-\omega_{k}\right)  $ so that we can approximate
\begin{equation}
P\left(  \omega-\omega_{k}\right)  \left\{  \mathbf{A}(\omega)\right\}
_{n_{T},n_{S}}\simeq P\left(  \omega-\omega_{k}\right)  \left\{
\mathbf{A}(\omega_{k})\right\}  _{n_{T},n_{S}}. \label{eq:traditional_approx}%
\end{equation}
In this situation, we can apply a constant precoder $\mathbf{A}(\omega_{k})$
to all the symbols that go through the $k$th subcarrier, which means that we
can in practice change the order of precoder and FBMC\ modulator with respect
to the ideal implementation (cf. Fig. \ref{fig:trad_MIMO_TXRX}). Under strong
frequency selectivity of the ideal precoder $\mathbf{A}(\omega)$, the
approximation in (\ref{eq:traditional_approx}) does not hold anymore and a
more accurate description of $\mathbf{A}(\omega)$ around $\omega_{k}$ is needed.

\begin{figure}[tbh]
\begin{center}
\includegraphics[width=\columnwidth]{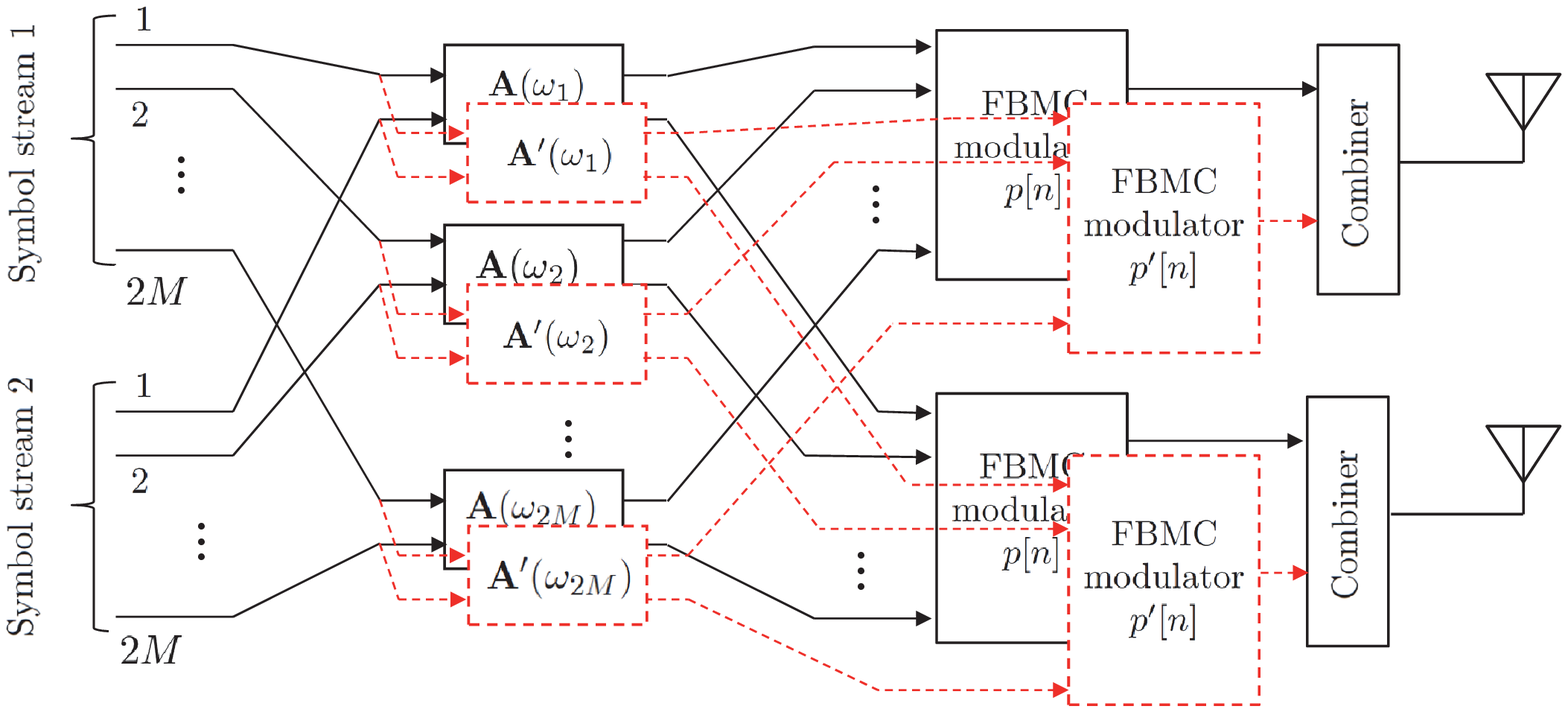}
\end{center}
\caption
{Proposed implementation of the frequecy-selective precoder for the specific case of $N_{T}%
=2$ transmit antennas and $N_{T}=2$ parallel stages. }
\label{fig:proposed_MIMO_TX}
\end{figure}

Assume that the entries of the precoding matrix $\mathbf{A}(\omega)$ are
analytic functions of $\omega$, so that they are expressible as their Taylor
series development around $\omega_{k}$, namely
\[
\mathbf{A}(\omega)=\sum_{\ell=0}^{\infty}\frac{1}{\ell!}\mathbf{A}^{(\ell
)}(\omega_{k})\left(  \omega-\omega_{k}\right)  ^{\ell}%
\]
where $\mathbf{A}^{(\ell)}(\omega_{k})$ denotes the $\ell$th derivative of
$\mathbf{A}(\omega)$\ evaluated at $\omega=\omega_{k}$. The idea behind the
classical precoder implementation in (\ref{eq:traditional_approx}) and in Fig.
\ref{fig:trad_MIMO_TXRX} is to truncate this Taylor series development and to
consider only its first term ($\ell=0$). Here, we suggest to go a bit further
and consider the truncation of the above series representation to include its
first $K_{T}$ terms ($T$ denoting transmit side), so that the transmitter
filter has an effective frequency response equal to
\begin{equation}
\mathbf{A}(\omega)\simeq\sum_{\ell=0}^{K_{T}-1}\frac{1}{\ell!}\left(
\omega-\omega_{k}\right)  ^{\ell}\mathbf{A}^{(\ell)}(\omega_{k}).
\label{eq:freq_response_truncated}%
\end{equation}
for $\omega$ sufficiently close to $\omega_{k}$. The main advantage of
extending this truncation to the case $K_{T}>1$ comes from the fact that one
can effectively implement the above filter by using $K_{T}$ parallel polyphase
FBMC modulators corresponding to the $K_{T}$ sums in
(\ref{eq:freq_response_truncated}), so that each of the $K_{T}$ parallel
precoders consists of single-tap per-subcarrier implementations. To see this,
assume that the prototype pulse $p[n]$ is a sampled version of an original
analog waveform $p(t)$ and define as $p^{(\ell)}[n]$ a sampled version of
$p^{(\ell)}(t)$, namely the $\ell$th derivative of the analog waveform $p(t)$.
These definitions will be more formally presented in Section
\ref{sec:performance_analysis}. Then, under certain regularity conditions on
the waveform $p(t)$, the Fourier transform of the sequence $p^{(\ell)}[n]$ can
be approximated for large $M$ as
\begin{equation}
P^{(\ell)}\left(  \omega\right)  \simeq\left(  2M\textnormal{j}%
\omega\right)  ^{\ell}P\left(  \omega\right)  . \label{eq:P_l}%
\end{equation}
Hence, by conveniently rewriting (\ref{eq:freq_response_truncated}), we see
that we can approximate
\begin{multline}
P\left(  \omega-\omega_{k}\right)  \mathbf{A}(\omega)\simeq\label{eq_A_KT}\\
\simeq\sum_{\ell=0}^{K_{T}-1}\frac{1}{\ell!}\left(  \frac{-\textnormal{j}%
}{2M}\right)  ^{\ell}P^{(\ell)}\left(  \omega-\omega_{k}\right)
\mathbf{A}^{(\ell)}(\omega_{k}).
\end{multline}
Now, observe that each term of the above sum has exactly the same form as the
first order (single-tap per-subcarrier) precoder $P\left(  \omega-\omega
_{k}\right)  \mathbf{A}(\omega_{k})$, replacing the actual precoder matrix
$\mathbf{A}(\omega_{k})$ and the original prototype pulse $P\left(
\omega\right)  $ by their derivative-associated counterparts $\mathbf{A}%
^{(\ell)}(\omega_{k})$, $P^{(\ell)}\left(  \omega\right)  $. From all this, we
can conclude that the $K_{T}$-term truncation of the ideal transmit precoder
frequency response can be generated by combining a set of $K_{T}$ parallel
conventional precoders. This is further illustrated in Fig.
\ref{fig:proposed_MIMO_TX}, where we represent the suggested implementation of
the transmit precoder when the number of parallel stages was fixed to
$K_{T}=2$ and the number of transmit antennas to $N_{T}=2$. We \ have
represented in red the additional stage that needs to be superposed to the
original one (in black), which is the same as in Fig. \ref{fig:trad_MIMO_TXRX}

\begin{figure}[tbh]
\begin{center}
\includegraphics[width=\columnwidth]{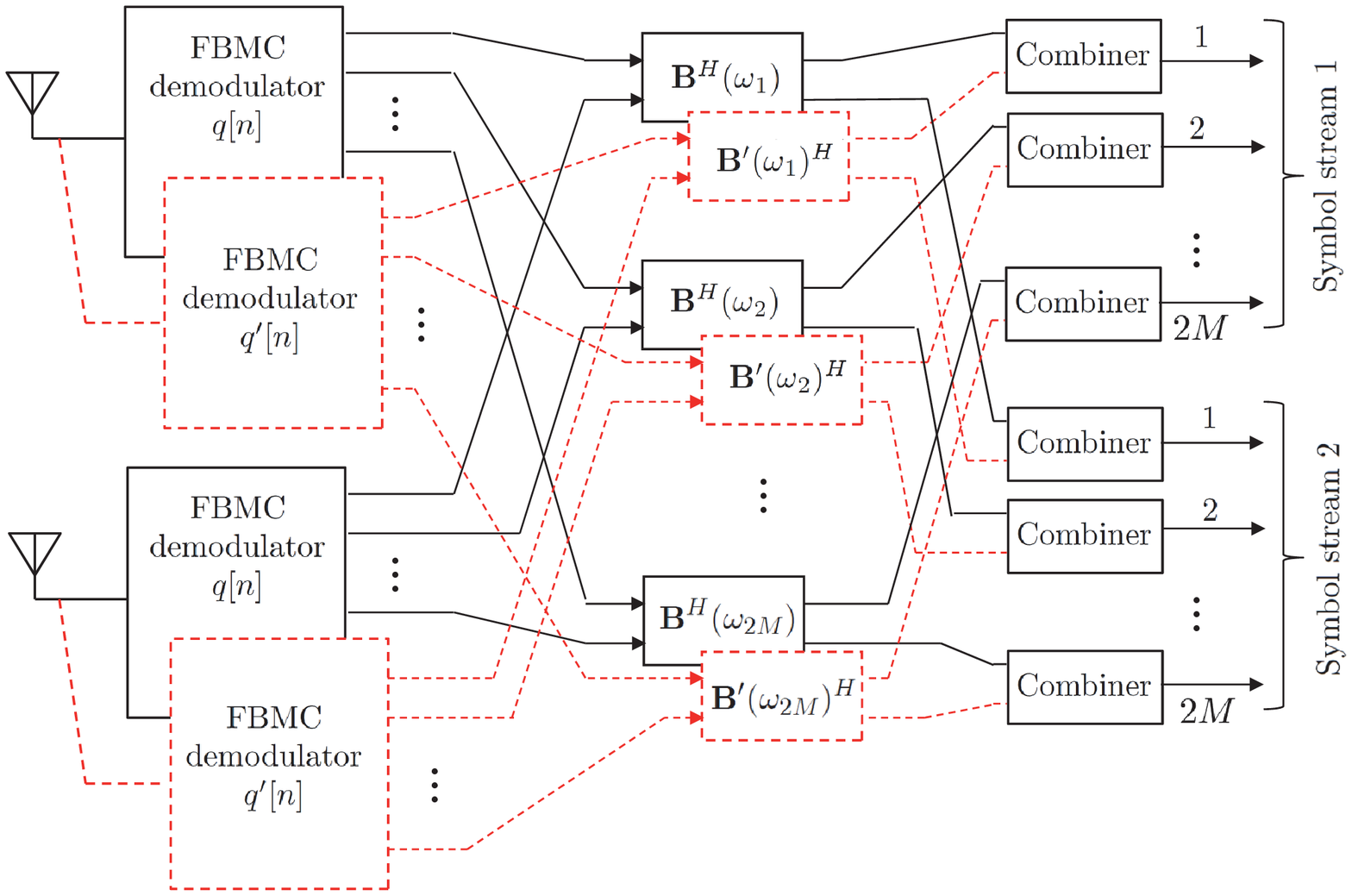}
\end{center}
\caption
{Proposed implementation of the frequecy-selective linear receiver for the specific case of $N_{R}%
=2$ receive antennas and $K_{R}=2$ parallel stages. }
\label{fig:proposed_MIMO_RX}
\end{figure}

We can follow the same approach in order to approximate the ideal
frequency-selective linear receiver matrix $\mathbf{B}(\omega)$ in combination
with the receive prototype pulse. Fig. \ref{fig:proposed_MIMO_RX} illustrates
the proposed architecture for the simple case of $N_{R}=2$ receive antennas
and $K_{R}=2$ parallel stages. From all the above, we can conclude that we can
approximate the ideal frequency-selective precoder/linear receiver as depicted
in Fig. \ref{fig:ideal_precoder_receiver} by simply increasing the number of
parallel stages ($K_{T}$, $K_{R}$) that are implemented at the transmitter and
at the receiver. In the following section we analyze the performance of the
proposed transceiver architecture in terms of the residual ISI/ICI distortion
at the output of the receiver.%

Let us now provide a more formal description of the proposed algorithm.
Consider the transmission of $N$ complex-valued (QAM) multicarrier symbols
through each of the $N_{S}$ MIMO streams. We will denote by $\mathbf{S}_{n}$,
$n=1\ldots N_{S}$, a $2M\times N$ matrix that contains, at each of its
columns, the multicarrier symbols that are transmitted through the $n$th
MIMO\ stream. Let $\mathcal{Z}_{n}$, $n=1\ldots N_{S}$, denote the
$2M\times\tilde{N}$ matrix of received samples corresponding to the reception
of the $n$th MIMO stream (see further Fig. \ref{fig:ideal_precoder_receiver}),
where $\tilde{N}$ is the total number of non-zero received samples associated
with the transmission of the $N$ multicarrier symbols (note that $\tilde
{N}\geq N$ due to the memory of the filterbank). Observe that $\mathcal{Z}%
_{n}$ will inherently have contributions from all the symbol matrices
$\mathbf{S}_{1},\ldots,\mathbf{S}_{N_{S}}$.

\begin{remark}
\label{remark_Delta}In what follows, given a general frequency-dependent
quantity $X=X(\omega)$ we define $\Lambda\left(  X\right)
=\textnormal{diag}\left\{  X(\omega_{1}),\ldots,X(\omega_{2M})\right\}  $,
i.e. a diagonal matrix containing the value of the function $X$ at the points
$\omega_{1},\ldots,\omega_{2M}$. We will also write $H_{n_{R},n_{T}}=\left\{
\mathbf{H}\left(  \omega\right)  \right\}  _{n_{R},n_{T}}$, $A_{n_{T},n_{_{S}%
}}^{\left(  \ell\right)  }=A_{n_{T},n_{_{S}}}^{\left(  \ell\right)  }%
(\omega)=\left\{  \mathbf{A}^{\left(  \ell\right)  }(\omega)\right\}
_{n_{T},n_{_{S}}}$ and $B_{n_{R},n}^{\left(  \ell\right)  }=B_{n_{R}%
,n}^{\left(  \ell\right)  }(\omega)=\left\{  \mathbf{B}^{\left(  \ell\right)
}(\omega)\right\}  _{n_{R},n}$. We will sometimes omit the dependence on
$\omega$ in several frequency-dependent quantities when this fact is clear
from the context.
\end{remark}

Let us first provide a formal description of the receive signal samples
$\mathcal{Z}_{n}$ associated with the $n$th MIMO symbol stream under the
traditional per-subcarrier design in Figure \ref{fig:trad_MIMO_TXRX}. The
$n_{S}$th symbol stream matrix $\mathbf{S}_{n_{S}}$ is precoded as
$\Lambda\left(  A_{n_{T},n_{_{S}}}\right)  \mathbf{S}_{n_{S}}$ for each of the
transmit antennas $n_{T}=1,\ldots,N_{T}$ and FBMC-modulated. The signal that
is transmitted through the $n_{T}$ antenna goes through the channel
$H_{n_{R},n_{T}}$ and is received by the $n_{T}$th antenna and
FBMC-demodulated. We will denote by
\begin{equation}
\mathcal{Z}_{p,q}^{H_{n_{R},n_{T}}}\left(  \Lambda\left(  A_{n_{T},n_{_{S}}%
}\right)  \mathbf{S}_{n_{S}}\right)  \label{eq:output_FBMCdemod}%
\end{equation}
the $2M\times\tilde{N}$ matrix of received samples at the output of the
corresponding FBMC demodulator, where $p,q$ are the prototype filters used at
the transmitter and the receiver respectively. In order to recover the
original symbols associated with the $n$th MIMO\ stream, a multiplicative
coefficient is finally applied at this signal at the per-subcarrier level, so
that the matrix in (\ref{eq:output_FBMCdemod}) is left multiplied by the
diagonal $\Lambda\left(  B_{n_{R},n}^{\ast}\right)  $. The total received
signal associated with the $n$th MIMO stream contains the contribution of all
transmit streams, all transmit and all receive antennas, so that it can be
expressed as%
\begin{equation}
\mathcal{Z}_{n}=\sum_{n_{R}=1}^{N_{R}}\sum_{n_{T}=1}^{N_{T}}\sum_{n_{S}%
=1}^{N_{s}}\Lambda\left(  B_{n_{R},n}^{\ast}\right)  \mathcal{Z}%
_{p,q}^{H_{n_{R},n_{T}}}\left(  \Lambda\left(  A_{n_{T},n_{_{S}}}\right)
\mathbf{S}_{n_{S}}\right)  \label{eq:Zn}%
\end{equation}
plus some additive noise that we omit in this discussion. Now, if the channel
and the precoder are sufficiently flat in the frequency domain, one may
approximate (see Appendix \ref{sec:appendix_theorem_proof}\ for a more formal
exposition)%
\begin{equation}
\mathcal{Z}_{p,q}^{H_{n_{R},n_{T}}}\left(  \Lambda\left(  A_{n_{T},n_{_{S}}%
}\right)  \mathbf{S}_{n_{S}}\right)  \simeq\Lambda\left(  H_{n_{R},n_{T}%
}A_{n_{T},n_{_{S}}}\right)  \mathcal{Y}_{p,q}\left(  \mathbf{S}_{n_{S}%
}\right)  \label{eq:defYpq}%
\end{equation}
where $\mathcal{Y}_{p,q}\left(  \mathbf{S}_{n_{S}}\right)  =\mathcal{Z}%
_{p,q}^{1}\left(  \mathbf{S}_{n_{S}}\right)  $ is the matrix of
FBMC-demodulated samples under an ideal SISO channel. Inserting this
approximation into (\ref{eq:Zn}) we see that
\[
\mathcal{Z}_{n}\simeq\sum_{n_{S}=1}^{N_{s}}\Lambda\left(  \left\{
\mathbf{B}^{H}\mathbf{HA}\right\}  _{n,n_{S}}\right)  \mathcal{Y}_{p,q}\left(
\mathbf{S}_{n_{S}}\right)  .
\]
If, additionally we force $\mathbf{B}^{H}\mathbf{HA=I}_{N_{S}}$, we will
approximately have $\mathcal{Z}_{n}\simeq\mathcal{Y}_{p,q}\left(
\mathbf{S}_{n}\right)  $ which will be a close approximation of the
transmitted symbols $\mathbf{S}_{n}$ if the prototype pulses are well designed.

Next, let us formulate the signal model under the proposed parallel
multi-stage architecture, assuming that the number of parallel stages is
$K_{T}$ at the transmit side and $K_{R}$ at the receive side. In order to
define the derivatives of the prototype pulses of (\ref{eq:P_l}) in a formal
manner, we make the following assumption:

$\mathbf{(As1)}$ The transmit and receive prototype pulses $p[n]$, $q[n]$ have
length $2M\kappa$, where $\kappa$ is the overlapping factor. Furthermore,
these pulses are obtained by discretization of smooth real-valued analog
waveforms $p(t)$, $q(t)$, which are smooth functions $\mathcal{C}^{R+1}\left(
\left[  -T_{s}\kappa/2,T_{s}\kappa/2\right]  \right)  $, $R\geq K_{T}+K_{R}$,
so that
\[
p[n]=p\left(  \left(  n-\frac{2M\kappa+1}{2}\right)  \frac{T_{s}}{2M}\right)
,\ n=1,\ldots,2M\kappa
\]
and equivalently for $q[n]$, where $T_{s}$ is the multicarrier symbol period.
Furthermore, the pulses $p(t)$, $q(t)$ and their $R+1$ sequential derivatives
are null at the end-points of their support, namely at $t=\pm T_{s}\kappa/2$.

Thanks to the above assumption, we can define $p^{(r)}$ and $\,q^{(r)}$ as the
sampled version of the $r$th derivative of $p(t)$ and $q(t)$ respectively,
that is
\[
p^{(r)}[n]=T_{s}^{r}p^{(r)}\left(  \left(  n-\frac{2M\kappa+1}{2}\right)
\frac{T_{s}}{2M}\right)  ,\ n=1,\ldots,2M\kappa
\]
and equivalently for $q^{(r)}$. In order to construct the proposed multi-stage
equalization system, we will assume that all quantities are sufficiently
smooth in the frequency domain, namely:

$\mathbf{(As2)}$ The frequency-depending quantities\ $\mathbf{A}%
(\omega),\mathbf{B}(\omega)$ and $\mathbf{H}(\omega)$ are $\mathcal{C}%
^{R^{\prime}+1}\left(  \mathbb{R}/2\pi\mathbb{Z}\right)  $ functions, where
$R^{\prime}>\left(  2R+1\right)  \left(  R+1\right)  $.
Furthermore\footnote{The following results can easily be generalized to the
case where $\mathbf{B}^{H}(\omega)\mathbf{H}(\omega)\mathbf{A}(\omega)$ is not
necessarily equal to the identity. However, we keep this assumption in order
to simplify the exposition.}, these matrices are constructed so that
$\mathbf{B}^{H}(\omega)\mathbf{H}(\omega)\mathbf{A}(\omega)=\mathbf{I}_{N_{s}%
}$.

Having established the definition of the time-domain derivative of the
prototype pulses and the smoothness conditions on precoder and channel, we can
now formulate the received signal model under the proposed parallel
multi-stage precoding/receiving architecture. Let $\mathcal{Z}_{n}^{\left(
\ell_{1},\ell_{2}\right)  }$ be defined as the $2M\times\tilde{N}$ received
signal matrix (equivalent to $\mathcal{Z}_{n}$), when transmitter and receiver
employ the $\ell_{1}$th and $\ell_{2}$th parallel stages respectively. Keeping
in mind that the $\ell$th parallel stage is constructed by replacing the
prototype pulse and the precoder/decoder by their corresponding $\ell$th order
derivatives, we can write
\begin{multline}
\mathcal{Z}_{n}^{\left(  \ell_{1},\ell_{2}\right)  }=\\
=\sum_{n_{_{S}}=1}^{N_{S}}\sum_{n_{R}=1}^{N_{R}}\sum_{n_{T}=1}^{N_{T}}%
\Lambda\left(  B_{n_{R},n}^{\left(  \ell_{2}\right)  \ast}\right)
\mathcal{Z}_{p^{(\ell_{1})},q^{(\ell_{2})}}^{H_{n_{R},n_{T}}}\left(
\Lambda\left(  A_{n_{T},n_{_{S}}}^{\left(  \ell_{1}\right)  }\right)
\mathbf{S}_{n_{S}}\right) \nonumber
\end{multline}
which is basically the same equation as (\ref{eq:Zn}), but replacing $\left\{
p,\mathbf{A}\right\}  $ by $\left\{  p^{(\ell_{1})},\mathbf{A}^{\left(
\ell_{1}\right)  }\right\}  $ and $\left\{  q,\mathbf{B}\right\}  $ by
$\left\{  q^{(\ell_{2})},\mathbf{B}^{(\ell_{2})}\right\}  $. The total
received signal is therefore described by the linear combination of the
signals that are transmitted and received by the multiple parallel stages,
using the coefficients established in (\ref{eq_A_KT}), namely
\begin{equation}
\mathcal{Z}_{n}=\sum_{\ell_{1}=0}^{K_{T}-1}\sum_{\ell_{2}=0}^{K_{R}-1}%
\frac{\left(  -\textnormal{j}\right)  ^{\ell_{1}+\ell_{2}}}{\ell_{1}%
!\ell_{2}!\left(  2M\right)  ^{\ell_{1}+\ell_{2}}}\mathcal{Z}_{n}^{\left(
\ell_{1},\ell_{2}\right)  }. \label{eq:receivedZ}%
\end{equation}
We claim that, assuming that precoder/receiver are constructed so that
$\mathbf{B}^{H}(\omega)\mathbf{H}(\omega)\mathbf{A}(\omega)\mathbf{=I}_{N_{S}%
}$, the above signal model is a very good approximation of the multicarrier
signal that would be received under frequency flat conditions, namely
$\mathcal{Y}_{p,q}\left(  \mathbf{S}_{n}\right)  $. This will be more formally
established in Section \ref{sec:performance_analysis}, where we provide an
asymptotic characterization of the resulting distortion error. In order to
provide these asymptotic results, we specifically focus on FBMC/OQAM
modulations, which allow perfect orthogonality conditions under an ideal channel.

\subsection{Specificities of the FBMC/OQAM signal model}

\begin{figure*}[tbh]
\begin{center}
\includegraphics[width=\columnwidth]{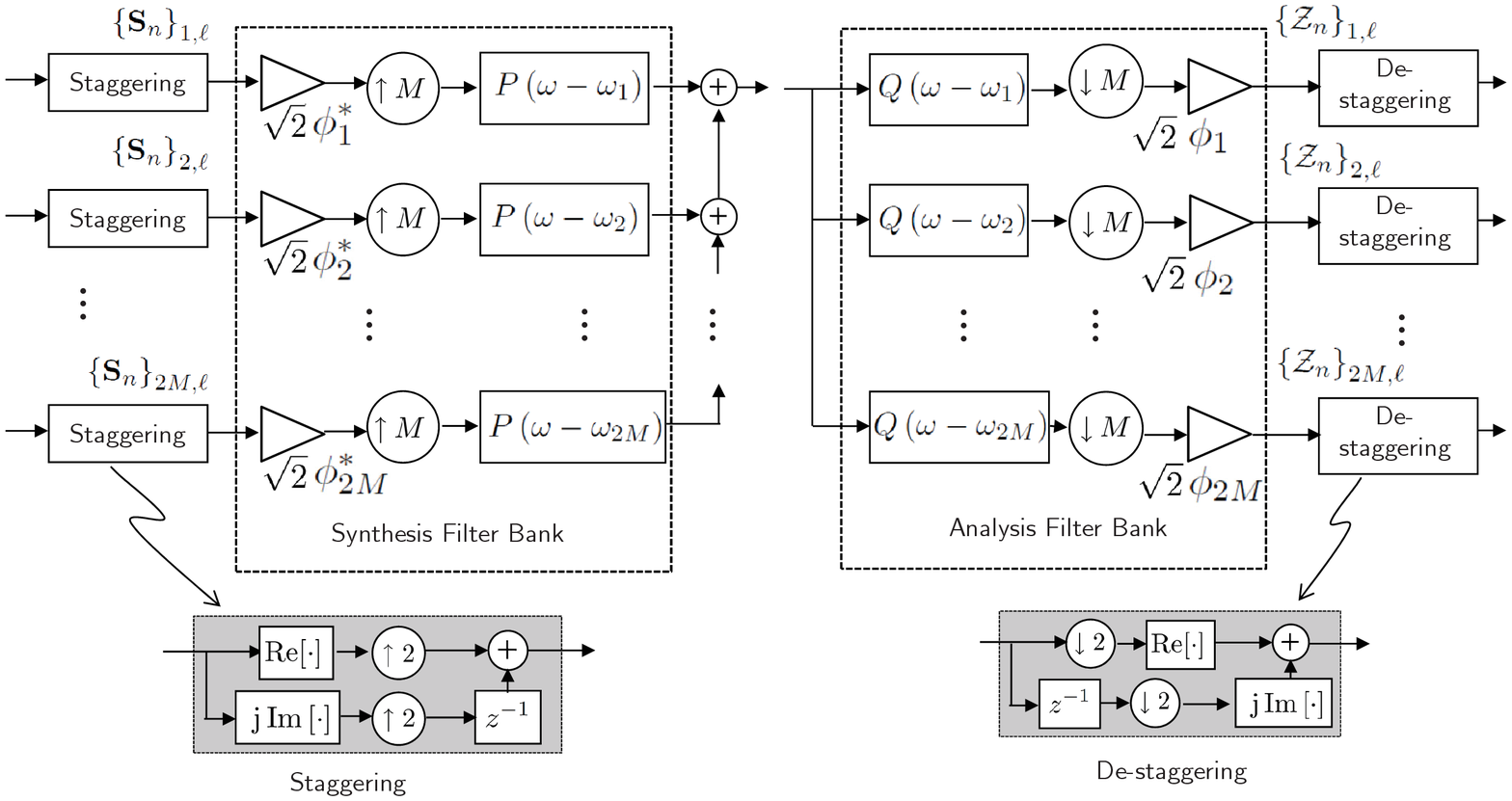}
\end{center}
\caption{Basic form of a FBMC/OQAM modulator and demodulator. Here, $P(\omega
)$ and $Q(\omega)$ are the transmit and receive prototype pulses, and $\phi
_k=\text{e}^{-\textnormal{j}\pi(M+1)(k-1)/(2M)}$}.
\label{fig:modFBMC}
\end{figure*}

The conceptual form of the FBMC/OQAM modulator and demodulator is illustrated
in Fig. \ref{fig:modFBMC}. As mentioned above, this modulation is widely
considered in the literature, thanks to the higher spectral efficiency with
respect to other filterbank multicarrier modulations and the possibility of
achieving perfect reconstruction of the transmitted symbols under perfect
channel conditions \cite{siohan02}. It can be described as a uniform,
critically sampled FBMC modulation scheme with different prototype pulses as
the transmitter ($p$) and the receiver ($q$), where the transmitted symbols
are drawn from a QAM\ modulation and staggered into an offset QAM\ (OQAM)
format.

As in the general description above, we consider that a total of $N$ complex
QAM\ multicarrier symbols ($2N$ real-valued symbols) are sequentially
transmitted through the $n$th stream, 
and let $\mathbf{S}_{n}$ denote a $2M\times 2N$ matrix that contains the symbols after the
staggering operation. Each pair of columns of $\mathbf{S}_{n}$ corresponds to a complex multicarrier symbol, and will be denoted by
$\mathbf{s}_{n}(\ell)$, $\ell=1,\ldots,N$. We will write $\mathbf{b}_{n}(\ell)=\textnormal{Re}\mathbf{s}_{n}(\ell)$ and $\mathbf{c}_{n}(\ell)=\textnormal{Im}
\mathbf{s}_{n}(\ell)$, and we will denote as $\mathbf{B}_n$ and $\mathbf{C}_n$ the $2M\times N$ matrices obtained by stacking these vectors in columns.
$\mathbf{b}_{n}(\ell)$ and $\mathbf{c}_{n}(\ell)$, so that $\mathbf{s}%
_{n}(\ell)=$ $\mathbf{b}_{n}(\ell)+\textnormal{j}\mathbf{c}_{n}(\ell)$. 
The signal matrix $\mathcal{Z}_{n}$ in (\ref{eq:receivedZ}) gathers the samples of
the received signal associated with the $n$th MIMO substream \emph{before the
de-staggering operation} (see Fig. \ref{fig:modFBMC}).

Following the notation in (\ref{eq:defYpq}), under an ideal SISO\ channel and
in the absence of precoder/receiver, the received samples matrix
$\mathcal{Z}_{n}$ corresponding to the complex symbols $\mathbf{S}_{n}$ will
be denoted by $\mathcal{Y}_{p,q}\left(  \mathbf{S}_{n}\right)  $. It can be
seen \cite{mestre13tsp} that matrix $\mathcal{Y}_{p,q}\left(  \mathbf{S}%
_{n}\right)  $ can be constrained to have dimensions \thinspace$2M\times
\left(  2N+2\kappa\right)  $. The number of columns of this matrix corresponds
to twice the number of transmitted multicarrier symbols ($2N$) plus some
additional columns ($2\kappa$) that account for the tail effects of the
prototypes $p,q$. From $\mathcal{Y}_{p,q}\left(  \mathbf{S}_{n}\right)  $, we
can construct two matrices $\mathcal{Y}_{p,q}^{\text{odd}}\left(
\mathbf{S}_{n}\right)  $, $\mathcal{Y}_{p,q}^{\text{even}}\left(
\mathbf{S}_{n}\right)  $ which contain its even- and odd-numbered columns, so
that
\begin{equation}
\mathcal{Y}_{p,q}\left(  \mathbf{S}_{n}\right)  =\mathcal{Y}_{p,q}%
^{\text{odd}}\left(  \mathbf{S}_{n}\right)  \otimes\lbrack0,1]+\mathcal{Y}%
_{p,q}^{\text{even}}\left(  \mathbf{S}_{n}\right)  \otimes\lbrack1,0]
\label{eq_Yodd_Yeven}%
\end{equation}
where $\otimes$ is the Kronecker product. According to the FBMC/OQAM
modulation, the original multicarrier symbols are retrieved by taking the
real/imaginary parts of the appropriate columns of $\mathcal{Y}_{n}%
^{\text{odd}}\left(  p,q\right)  $ and $\mathcal{Y}_{n}^{\text{even}}\left(
p,q\right)  $, that is via a de-staggering operation%
\begin{equation}
\mathbf{\hat{s}}_{n}(\ell)=\textnormal{Re}\left\{  \mathcal{Y}%
_{p,q}^{\text{odd}}\left(  \mathbf{S}_{n}\right)  \right\}  _{:,\ell+\kappa
-1}+\textnormal{j}\textnormal{Im}\left\{  \mathcal{Y}_{p,q}^{\text{even}%
}\left(  \mathbf{S}_{n}\right)  \right\}  _{:,\ell+\kappa}\text{ .}
\label{eq_ahat_OQAM}%
\end{equation}
A exact expression of $\mathcal{Y}_{p,q}^{\text{odd}}\left(  \mathbf{S}%
_{n}\right)  $ and $\mathcal{Y}_{p,q}^{\text{even}}\left(  \mathbf{S}%
_{n}\right)  $ can be found in \cite[(3)-(4)]{mestre13tsp}, see further
(\ref{eq_expr_Yeven}) in Appendix \ref{sec:appendix_theorem_proof}.

It is well known \cite{siohan02} that one can choose $p,q$ to meet some
\textquotedblleft bi-orthogonality\textquotedblright\ or perfect
reconstruction (PR) conditions, which guarantee that $\mathbf{\hat{s}}_{n_{S}%
}(\ell)=\mathbf{s}_{n_{S}}(\ell)$ in (\ref{eq_ahat_OQAM}). In order to
formulate these conditions, let $\mathbf{P}$ and $\mathbf{Q}$ denote two
$2M\times\kappa$ matrices obtained by arranging the original prototype pulse
samples in columns. In other words, the $k$th row of $\mathbf{P}$ (resp.
$\mathbf{Q}$) contains the $k$th polyphase component of the original pulse $p$
(resp. $q$). Next, consider two $2M\times\left(  2\kappa-1\right)  $ matrices
$\mathcal{R}\left(  p,q\right)  $ and $\mathcal{S}\left(  p,q\right)  $
obtained as
\begin{align}
\mathcal{R}\left(  p,q\right)   &  =\mathbf{P\circledast J}_{2M}%
\mathbf{Q}\label{def_R}\\
\mathcal{S}\left(  p,q\right)   &  =\left(  \mathbf{J}_{2}\otimes
\mathbf{I}_{M}\right)  \mathbf{P\circledast J}_{2M}\mathbf{Q} \label{def_S}%
\end{align}
where $\mathbf{\circledast}$ indicates row-wise convolution between matrices
and $\mathbf{J}_{2M}$ is the anti-identity matrix of size $2M$. It is well
known that one can impose PR conditions on the pulses $p$ and $q$ by imposing
\cite{mestre13tsp}%
\begin{equation}
\mathbf{U}^{+}\mathcal{R}\left(  p,q\right)  =\mathcal{I},\quad\mathbf{U}%
^{-}\mathcal{S}\left(  p,q\right)  =\mathbf{0} \label{eq_PRconds}%
\end{equation}
where $\mathbf{U}^{+}=\mathbf{I}_{2}\otimes\left(  \mathbf{I}_{M}%
+\mathbf{J}_{M}\right)  $, $\mathbf{U}^{-}=\mathbf{I}_{2}\otimes\left(
\mathbf{I}_{M}-\mathbf{J}_{M}\right)  $, and where $\mathcal{I}$ is a
$2M\times\left(  2\kappa-1\right)  $ matrix with ones in its central column
and zeros elsewhere. The above PR conditions can be significantly simplified
when $p=q$ and assuming that the prototype pulses are symmetric or
anti-symmetric in the time domain \cite{siohan02}.

\section{\label{sec:performance_analysis}Performance analysis under FBMC/OQAM}

Ideally, one would like to have $\mathcal{Z}_{n}$ as similar as possible to
the signal of the output of the decimators in the FBMC demodulators when the
ideal frequency-selective precoder and linear receiver are used (Fig.
\ref{fig:ideal_precoder_receiver}). In practice, however this only holds
approximately, in the sense that $\mathcal{Z}_{n}$ in (\ref{eq:receivedZ}%
)\ can be written as \
\begin{equation}
\mathcal{Z}_{n}=\mathcal{Y}_{p,q}\left(  \mathbf{S}_{n}\right)  +\mathcal{E}%
_{p,q}\left(  \mathbf{S}_{n}\right)  \label{eq_Sndecomposed}%
\end{equation}
for some error $\mathcal{E}_{p,q}\left(  \mathbf{S}_{n}\right)  $. More
specifically, decomposing $\mathcal{Z}_{n}$ into $\mathcal{Z}_{n}^{\text{odd}%
}$ and $\mathcal{Z}_{n}^{\text{even}}$ as in (\ref{eq_Yodd_Yeven}), one would
estimate the $\ell$th multicarrier symbol as%
\begin{equation}
\mathbf{\hat{s}}_{n}(\ell)=\textnormal{Re}\left\{  \mathcal{Z}%
_{n}^{\text{odd}}\right\}  _{:,\ell+\kappa-1}+\textnormal{j}%
\textnormal{Im}\left\{  \mathcal{Z}_{n}^{\text{even}}\right\}
_{:,\ell+\kappa}\text{ .} \label{eq:def_ahat}%
\end{equation}
Two different sources of error will be present in this estimation of the
symbols: an implementation error due to the frequency selectivity of the
system, namely $\mathcal{E}_{p,q}\left(  \mathbf{S}_{n_{S}}\right)  $, plus a
representation error which arises from the fact that the PR conditions in
(\ref{eq_PRconds}) may not hold. In this section, we characterize the behavior
of the total resulting error by assuming that the number of subcarriers is
asymptotically large ($M\rightarrow\infty$). The following additional
assumptions will be needed in order to provide the corresponding result:

$\mathbf{(As3)}$ The transmitted complex symbols are drawn from a bounded constellation.

$(\mathbf{As4})$\ The real and imaginary parts of the transmitted symbols are
independent, identically distributed random variables with zero mean and power
$P_{s}/2$.

Under these assumptions, it is possible to characterize the behavior of the
residual distortion at the output of the receiver, assuming that the number of
subcarriers is asymptotically high ($M\rightarrow\infty$). In order to
formulate the result, we need some additional definitions, that are presented
next. Given four integers $m,n,m^{\prime},n^{\prime}$, we define
$\eta_{(m,n,m^{\prime},n^{\prime})}^{\left(  +,-\right)  }$ as the following
pulse-specific quantity:
\begin{align*}
\eta_{(m,n,m^{\prime},n^{\prime})}^{\left(  +,-\right)  }  &  =\frac{P_{s}%
}{2M}\textnormal{tr}%
\Bigg[%
\mathcal{R}\left(  p^{(m)},q^{(n)}\right)  \mathcal{R}^{T}\left(
p^{(m^{\prime})},q^{(n^{\prime})}\right)  \mathbf{U}^{+}\\
&  +\mathcal{S}\left(  p^{(m)},q^{(n)}\right)  \mathcal{S}^{T}\left(
p^{(m^{\prime})},q^{(n^{\prime})}\right)  \mathbf{U}^{-}%
\Bigg]%
\end{align*}
where $\mathcal{R(}$\textperiodcentered$,$\textperiodcentered$)$ and
$\mathcal{S(}$\textperiodcentered$,$\textperiodcentered$)$ are defined in
(\ref{def_R})-(\ref{def_S}). The quantity $\eta_{(m,n)}^{\left(  -,+\right)
}$ is equivalently defined, but swapping $\mathbf{U}^{+}$ and $\mathbf{U}^{-}$
in the above equation. Let $\Psi_{K}^{\left(  +,-\right)  }$ denote a
$2\times2$ matrix constructed as%
\begin{equation}
\Psi_{K}^{\left(  +,-\right)  }=\left[
\begin{array}
[c]{cc}%
\eta_{(K,0,K,0)}^{\left(  +,-\right)  }\mathbb{I}_{\left\{  K_{T}=K\right\}  }
& \eta_{(K,0,0,K)}^{\left(  +,-\right)  }\mathbb{I}_{\left\{  K_{R}%
=K_{T}\right\}  }\\
\eta_{(K,0,0,K)}^{\left(  +,-\right)  }\mathbb{I}_{\left\{  K_{R}%
=K_{T}\right\}  } & \eta_{(0,K,0,K)}^{\left(  +,-\right)  }\mathbb{I}%
_{\left\{  K_{R}=K\right\}  }%
\end{array}
\right]  \label{eq_def_PsiK}%
\end{equation}
where $\mathbb{I}_{\left\{  \text{\textperiodcentered}\right\}  }$ is the
indicator function. Let $\Psi_{K}^{\left(  -,+\right)  }$ be constructed as
$\Psi_{K}^{\left(  +,-\right)  }$ but changing all instances of $\left(
+,-\right)  $ for $\left(  -,+\right)  $. \ The following quantities will take
into account the fact that PR conditions may not hold
\begin{align*}
\delta &  =\frac{P_{s}}{2M}\textnormal{tr}%
\Bigg[%
\left(  \mathcal{R}\left(  p,q\right)  -\frac{1}{2}\mathcal{I}\right)  \left(
\mathcal{R}\left(  p,q\right)  -\frac{1}{2}\mathcal{I}\right)  ^{T}%
\mathbf{U}^{+}\\
&  +\mathcal{S}\left(  p,q\right)  \mathcal{S}^{T}\left(  p,q\right)
\mathbf{U}^{-}%
\Bigg]%
\end{align*}%
\begin{align*}
\mu_{(m,n)}  &  =\frac{P_{s}}{2M}\textnormal{tr}%
\Bigg[%
\left(  \mathcal{R}\left(  p,q\right)  -\frac{1}{2}\mathcal{I}\right)
\mathcal{R}^{T}\left(  p^{(m)},q^{(n)}\right)  \mathbf{U}^{+}\\
&  +\mathcal{S}\left(  p,q\right)  \mathcal{S}^{T}\left(  p^{(m)}%
,q^{(n)}\right)  \mathbf{U}^{-}%
\Bigg]%
\\
\tilde{\mu}_{\left(  \ell,m\right)  }^{(K)}  &  =\sum_{j=K}^{\ell}\left(
-1\right)  ^{j+K}\binom{\ell}{j}\binom{j-1}{K-1}\mu_{\left(  j,m-j\right)  }.
\end{align*}
Indeed, observe that these two quantities are zero under the PR conditions in
(\ref{eq_PRconds}): clearly $\mathbf{U}^{-}\mathcal{S}\left(  p,q\right)
=\mathbf{0}$ whereas
\[
\mathbf{U}^{+}\left(  \mathcal{R}\left(  p,q\right)  -\frac{1}{2}%
\mathcal{I}\right)  =\mathbf{U}^{+}\mathcal{R}\left(  p,q\right)
-\mathcal{I}=\mathbf{0.}%
\]
We will additionally need some channel-specific functions $\alpha_{n,n_{S}%
}^{(m,\ell)}\left(  \omega\right)  $, $\beta_{n,n_{S}}^{(m,\ell)}\left(  \omega\right)
$ and $\gamma_{n,n_{S}}\left(  \omega\right)  $, defined as
\begin{align}
\alpha_{n,n_{S}}^{(m,\ell)}  &  =\frac{\sqrt{2}\left(  -\textnormal{j}%
\right)  ^{m}}{m!}\binom{m}{\ell}\left\{  \left(  \mathbf{B}^{H}%
\mathbf{H}\right)  ^{\left(  m-\ell\right)  }\mathbf{A}^{\left(  \ell\right)
}\right\}  _{n,n_{S}}\label{eq:def_alpha}\\
\beta_{n,n_{S}}^{(m,\ell)}  &  =\frac{\sqrt{2}\left(  -\textnormal{j}%
\right)  ^{m}}{m!}\binom{m}{\ell}\left\{  \mathbf{B}^{\left(  \ell\right)
H}\left(  \mathbf{HA}\right)  ^{\left(  m-\ell\right)  }\right\}  _{n,n_{S}%
}\label{eq:def_beta}\\
\gamma_{n,n_{S}}  &  =\frac{\sqrt{2}\left(  -\textnormal{j}\right)
^{K_{R}+K_{T}}}{K_{T}!K_{R}!}\left\{  \mathbf{B}^{\left(  K_{R}\right)
H}\mathbf{HA}^{\left(  K_{T}\right)  }\right\}  _{n,n_{S}}.
\label{eq:def_gamma}%
\end{align}
where we have omitted the dependence of all quantities on $\omega$ to ease the
notation and where $\ell\leq m$. We have now all the ingredients to
characterize the asymptotic distortion power associated with the $n$th
parallel symbol stream observed at the $k$th subcarrier output of the
FBMC/OQAM demodulator, which will be denoted by $P_{e}\left(  k,n\right)  $,
$1\leq k\leq2M$, $1\leq n\leq N_{S}$.

\begin{theorem}
\label{th:Pe}Consider the linear parallelized MIMO FBMC system presented
above, with $K_{T}\geq1$ parallel stages at the transmitter and $K_{R}\geq1$
parallel stages at the receiver. Let $\mathbf{\hat{s}}_{n}\left(  \ell\right)
$ be as defined in (\ref{eq:def_ahat}), i.e. as the estimate of $\mathbf{s}%
_{n}\left(  \ell\right)  $, the $\ell$th column vector of the complex-valued
symbol matrix. Assume that $\mathbf{(As1)-(As4)}$ hold and let $K=\min\left(
K_{T},K_{R}\right)  $. Then for any $\ell\in\left\{  \kappa,\ldots
,N-\kappa\right\}  $ one can write%
\[
\mathbb{E}\left[  \left\vert \left\{  \mathbf{\hat{s}}_{n}\left(  \ell\right)
\right\}  _{k}-\left\{  \mathbf{s}_{n}\left(  \ell\right)  \right\}
_{k}\right\vert ^{2}\right]  =P_{e}\left(  k,n\right)  +o\left(
M^{-2K}\right)
\]
as $M\rightarrow\infty$. The term $P_{e}\left(  k,n\right)  $ can be
decomposed in two terms, namely $P_{e}\left(  k,n\right)  =P_{e,1}\left(
k,n\right)  +P_{e,2}\left(  k,n\right)  $, with
\begin{align*}
&  P_{e,1}\left(  k,n\right)  =2\delta\\
&  -\sum_{m=K_{R}}^{2K}\frac{2\sqrt{2}}{\left(  2M\right)  ^{m}}\sum
_{\ell=K_{R}}^{m}\mu_{\left(  0,m\right)  }\textnormal{Re}\left(
\beta_{n,n}^{(m,\ell)}\right) \\
&  -\sum_{m=K_{T}}^{2K}\frac{2\sqrt{2}}{\left(  2M\right)  ^{m}}\sum
_{\ell=K_{T}}^{m}\tilde{\mu}_{\left(  \ell,m\right)  }^{(K_{T})}%
\textnormal{Re}\left(  \alpha_{n,n}^{(m,\ell)}\right) \\
&  +\frac{2\sqrt{2}}{\left(  2M\right)  ^{2K}}\textnormal{Re}\gamma_{n,n}%
\mu_{(K,K)}\mathbb{I}_{\left\{  K_{R}=K_{T}\right\}  }%
\end{align*}
and
\begin{align*}
P_{e,2}\left(  k,n\right)   &  =\frac{1}{\left(  2M\right)  ^{2K}}%
\sum_{n_{_{S}}=1}^{N_{S}}\textnormal{Re}^{T}\left[  \xi_{n,n_{S}}%
^{(K,K)}\right]  \Psi_{K}^{\left(  +,-\right)  }\textnormal{Re}\left[
\xi_{n,n_{S}}^{(K,K)}\right] \\
&  +\frac{1}{\left(  2M\right)  ^{2K}}\sum_{n_{_{S}}=1}^{N_{S}}%
\textnormal{Im}^{T}\left[  \xi_{n,n_{S}}^{(K,K)}\right]  \Psi_{K}^{\left(
-,+\right)  }\textnormal{Im}\left[  \xi_{n,n_{S}}^{(K,K)}\right]
\end{align*}
where $\xi_{n,n_{S}}^{(m,\ell)}\left(  \omega\right)  =\left[  \alpha_{n,n_{S}%
}^{(m,\ell)}\left(  \omega\right)  ,\beta_{n,n_{S}}^{(m,\ell)}\left(  \omega\right)
\right]  ^{T}$ and where all the frequency-dependent quantities ($\alpha
_{n,n_{S}}^{(m,\ell)},\beta_{n,n_{S}}^{(m,\ell)},\gamma_{n,n}$) are evaluated at
$\omega=\omega_{k}$.
\end{theorem}

\begin{IEEEproof}
See Appendix \ref{sec:appendix_theorem_proof}.
\end{IEEEproof}

According to the above result, the inherent distortion of the FBMC/OQAM
modulation can be asymptotically decomposed into two terms, $P_{e,1}\left(
k,n\right)  $ and $P_{e,2}\left(  k,n\right)  $. The first term basically
accounts for the fact that the prototype pulses $p,q$ need not have PR
conditions. It can readily be observed that this term is identically zero when
the conditions in (\ref{eq_PRconds}) hold. The second term $P_{e,2}\left(
k,n\right)  $ inherently describes the effect of the residual distortion
caused by the channel frequency selectivity, even when PR\ conditions hold.
This term essentially decays as $O\left(  M^{-2K}\right)  $ when
$M\rightarrow\infty$, where $K$ is the minimum between transmit and receive
parallel stages. This means that if both the\ transmit and the\ receive
processing matrices are frequency-selective, it does not make much sense to
increase the number of parallel stages at one side of the communications link
beyond the number of stages at the other, since the asymptotic behavior will
ultimately be dictated by the minimum between the two. The situation is
different when only one of the matrices (either $\mathbf{A}(\omega)$ or
$\mathbf{B}(\omega)$) is frequency-selective. In this case, the frequency flat
matrix can be seen as its exact representation in Taylor series, which is
equivalent to stating that the matrix is approximated using an infinite number
of terms (most of which are zero), i.e. $K_{T}=\infty$ or $K_{R}=\infty$. In
this situation, increasing the number of stages that implement the
frequency-selective matrix will always have a beneficial effect.

On the other hand, one should also observe from the expression of
$P_{e}\left(  k,n\right)  $ that the total residual distortion power that is
observed at the $n$th receive symbol stream is an additive combination of the
distortion associated with each of the transmit symbol streams (note the sum
from $n_{S}=1$ to $N_{S}$ in the asymptotic expression for $P_{e,2}\left(
k,n\right)  $). This justifies the claim that general MIMO processing is very
vulnerable to the presence of highly frequency-selective channels, since the
higher the number of parallel streams, the higher the residual distortion
power that will be observed at the output of the receiver. Furthermore, the
expression of $P_{e}\left(  k,n\right)  $ provides a very convenient way of
fixing the number of parallel stages at the transmitter and receiver ($K$) in
order to guarantee a certain degree of performance. Given a triplet of
channel, precoder and linear receiver ($\mathbf{A}(\omega)$, $\mathbf{H}%
(\omega)$ and $\mathbf{B}(\omega)$) one only needs to evaluate $P_{e}\left(
k,n\right)  \,$\ in order to obtain the minimum $K$ that guarantees a
sufficiently low distortion power.

Finally, it is worth pointing out that the asymptotic residual distortion
expression presented in Theorem \ref{th:Pe} generalizes the one obtained in
\cite{mestre13tsp} for SISO channels in different important aspects. Here,
both transmit and receive frequency-selective processing structures are
considered, whereas only receive processing (equalization) was assumed in
\cite{mestre13tsp}. Furthermore, the\ above expression of $P_{e}\left(
k,n\right)  $ above does not assume PR conditions on the prototype pulses,
which was not the case in \cite{mestre13tsp}. Section \ref{sec:numerical}
shows that this asymptotic expression provides an extremely accurate
description of the system behavior under severe channel frequency selectivity.

\subsection{Computational Complexity and Latency}

Contrary to multi-tap filter-based solutions that process the signal per
subcarrier using a finite impulse response (FIR)\ filter, the proposed
parallel multi-stage architecture incurs in no additional penalty in terms of
latency. This is because all the constituent stages can be implemented in
parallel, avoiding all the unnecessary delays of other multi-tap based
filtering approaches. Note that the insertion of a multi-tap processor per
subcarrier will generally incur in a latency increase proportional to the
product between the number of taps and the number of subcarriers, which may
not be tolerable in delay-critical applications.

As for the associated complexity of the proposed multi-stage architecture, we
can evaluate it in terms of the total number of real-valued multiplications
and sums. We will consider a transmit/receive filterbank implementation using
an FFT-based polyphase architecture\ \cite{siohan02}, assuming that the number
of subcarriers is a power of $2$ and that the prototype pulses are symmetric
in the time domain. Using the split-radix algorithm, one can implement an FFT
by only using $2M\left(  \log_{2}M-1\right)  +4$ real-valued multiplications
and $6M\log_{2}M+4$ real valued sums \cite{duhamel90}. Using this together
with the fact that the prototype pulse is real-valued and that each complex
product can be implemented with $3$ real-valued multiplication plus $5$
real-valued sums, we can establish the total number of real-valued sums and
multiplications of the multi-stage architecture given in Table
\ref{table_comp}. In this table, we have disregarded the terms of order $o(M)$
and have also introduced the complexity of a MIMO multi-tap equalizer
\cite{Ihalainen2011} for comparison purposes. These numbers will be used in
the numerical analysis of the following section.%

\begin{table}[tbp] \centering

\begin{tabular}
[c]{|l|l|}\hline
\textbf{Algorithm} & \textbf{Real-valued products}\\\hline
Multi-stage (TX) & $2MK_{T}\left[  N_{T}\log_{2}M+\left(  \kappa+2\right)
N_{T}+2N_{S}N_{T}\right]  $\\
Multi-stage (RX) & $2MK_{R}\left[  N_{R}\log_{2}M+\left(  \kappa+2\right)
N_{R}+3N_{R}N_{S}\right]  $\\
Multi-tap (RX) & $2M\left[  N_{R}\log_{2}M+\left(  \kappa+2\right)
N_{R}+3N_{S}N_{R}\left(  N_{taps}+1\right)  \right]  $\\\hline
\textbf{Algorithm} & \textbf{Real-valued sums}\\\hline
Multi-stage (TX) & $2MK_{T}\left[  3N_{T}\log_{2}M+\left(  2\kappa+1\right)
N_{T}+2N_{S}N_{T}\right]  $\\\hline
Multi-stage (RX) & $2MK_{R}\left[  3N_{R}\log_{2}M+\left(  2\kappa+3\right)
N_{R}+\left(  7N_{R}-2\right)  N_{S}\right]  $\\\hline
Multi-tap (RX) & $2M[3N_{R}\log_{2}M+\left(  2\kappa+3\right)  N_{R}+\left(
7N_{R}-2\right)  N_{SS}$\\
& $+\left(  \left(  7N_{R}-5\right)  N_{taps}-2\right)  N]$\\\hline
\end{tabular}
\caption{Total number of real-valued sums and multiplications associated with the parallel multi-stage architecture and a multi-tap MIMO equalizer with $N_{taps}$ matrix coefficients. \label{table_comp}}
\end{table}

\section{\label{sec:numerical}Numerical Analysis}

In this section, we analyze the performance of the proposed precoding/linear
receiver architectures in an LTE-like FBMC/OQAM system with an intercarrier
separation of $15$kHz and QPSK\ modulated symbols. We will assume that the
channel state information is perfectly known at the receiver, and also at the
transmitter whenever the use of frequency-selective processing is considered.
As for the actual FBMC modulation, we consider the PHYDYAS non-perfect
reconstruction (NPR)\ prototype pulse \cite{bellanger01, viholainen09} with
overlapping factor equal to $\kappa=3$. The same prototype pulse is used at
both transmitter and receiver. All MIMO\ channels were simulated as
independent, static and frequency-selective with a power delay profile given
by the ITU\ Extended Vehicular A (EVA) and Extended Typical Urban (ETU) models
\cite{36.1012013}.

\subsection{Validation of the asymptotic ICI/ISI\ distortion expressions}

\begin{figure}[tbh]
\begin{center}
\includegraphics[width=9cm]{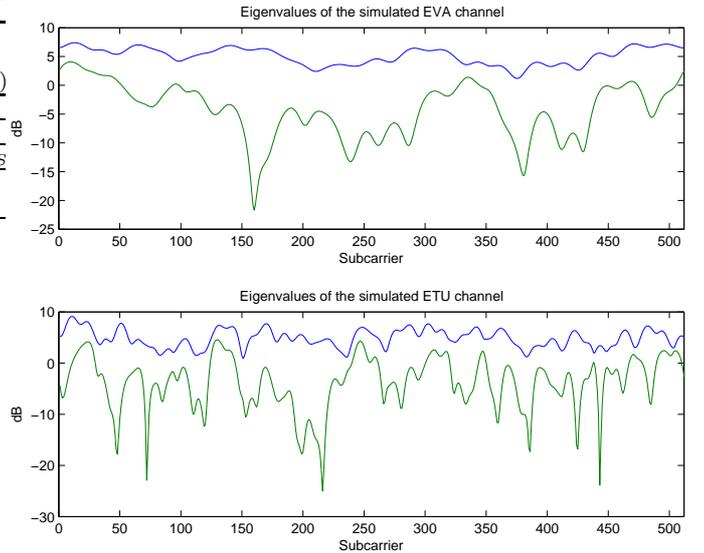}
\end{center}
\caption{Eigenvalues of the MIMO channels used\ in the first part of the
simulations, drawn from the EVA and ETU\ channel models. }%
\label{fig:evaetu}%
\end{figure}

In order to validate the expressions for the residual ICI/ISI distortion
provided in Theorem \ref{th:Pe}, we considered a noiseless scenario with $512$
subcarriers and two fixed channel impulse responses drawn from the EVA and the
ETU\ channel models. The number of antennas was fixed to $2$ at both the
transmitter and the receiver, namely $N_{T}=N_{R}=2$, and two different symbol
streams were transmitted $N_{S}=2$. Fig. \ref{fig:evaetu} shows the
eigenvalues of the simulated channel in the frequency domain. A set of $10000$
multicarrier symbols was randomly drawn from a QPSK\ modulation and the
corresponding signal to distortion power ratio was measured at the output of
the receiver. The simulated transceiver consisted of an eigenvector-based
precoder, where $\mathbf{A}(\omega)$ was selected as the dominant eigenvectors
of $\mathbf{H}^{H}(\omega)\mathbf{H}(\omega)$ and where $\mathbf{B}(\omega)$
inverted the resulting channel. 

\begin{figure}[tbh]
\begin{center}
\includegraphics[width=9cm]{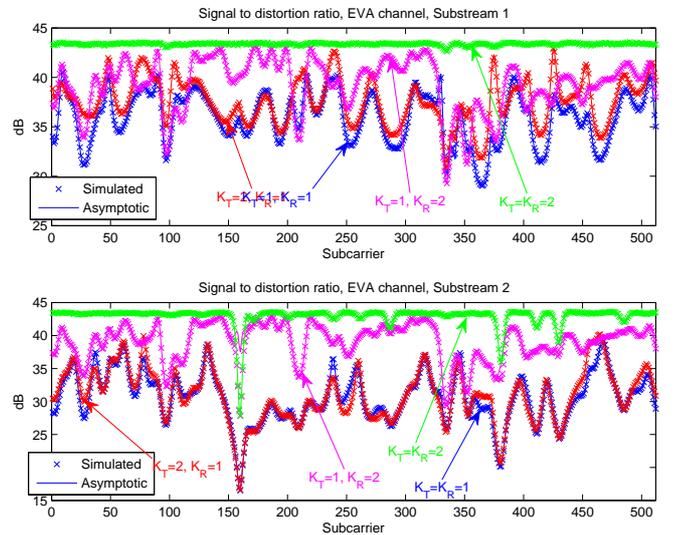}
\end{center}
\caption{Signal to distortion power ratio measured at the output of the
receiver when the transmitter uses SVD-type precoding and the receiver
performs channel inversion. The simulated channel was the one represented in
the upper plot of Fig. \ref{fig:evaetu}.}%
\label{fig:svd_eva}%
\end{figure}

\begin{figure}[tbhtbh]
\begin{center}
\includegraphics[width=9cm]{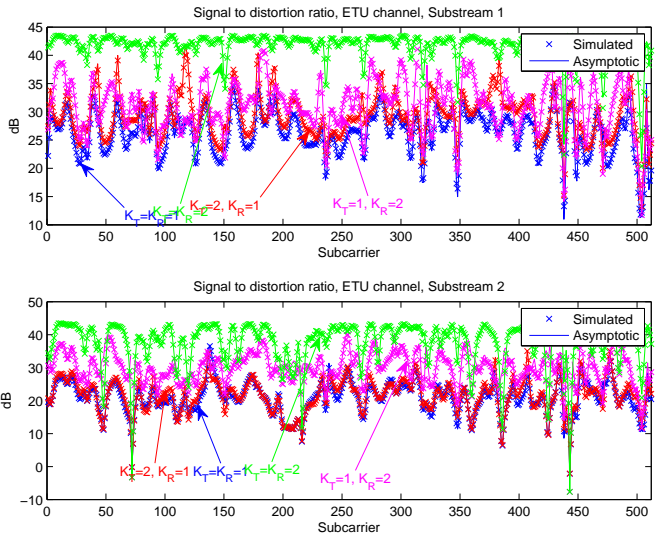}
\end{center}
\caption{Signal to distortion power ratio measured at the output of the
receiver when the transmitter uses SVD-type precoding and the receiver
performs channel inversion. The simulated channel was the one represented in
the lower plot of Fig. \ref{fig:evaetu}.}%
\label{fig:svd_etu}%
\end{figure}

Figs. \ref{fig:svd_eva} and \ref{fig:svd_etu} compare the simulated and
asymptotic performance as predicted by Theorem \ref{th:Pe} for different
values of the number of parallel stages at the transmitter/receiver, i.e.
$K_{T}$, $K_{R}$. In these two figures, solid lines represent the theoretical
performance as described by $P_{e}\left(  k,n\right)  $ whereas cross markers
are simulated performance values. Observe that there is a perfect match
between them, and the\ simulated results are virtually indistinguishable from
the theoretical ones, even for relatively moderate values of $M$. The only
rare\ differences between simulated and asymptotic performance become apparent
in situations where the coefficient of the second order term becomes
substantially high and the first order characterization so that the first
order fails to capture the actual distortion behavior.

As for the actual performance of the multi-stage transceiver architecture, it
is clearly seen that substantial gains can be achieved in terms of residual
ICI/ISI\ reduction by simply implementing a second parallel stage at the
transmitter/receiver. On the other hand, simulations confirm the fact that the
performance is roughly dictated by the minimum number of parallel stages used
at the transmit and receive sides, that is the minimum between $K_{R}$ and
$K_{T}$. In other words, when using frequency-selective processing at both
transmitter and receiver, the most important gains can be obtained by
considering the proposed architecture at both sides of the communications
link, but using the same number of parallel stages.

\subsection{Performance under general frequency-selective channels}

\begin{figure}[tbh]
\begin{center}
\includegraphics[width=9cm]{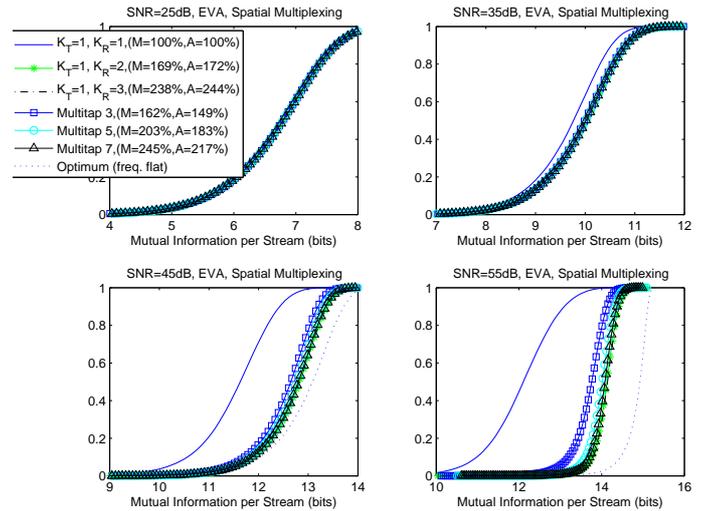}
\end{center}
\caption{Mutual information distribution (bits per stream) for different
levels of the background noise with $N_{T}=2$, $N_{R}=4$ and spatial
multiplexing (EVA\ channel model). }%
\label{fig:eva_2x4}%
\end{figure}

\begin{figure}[tbhtbh]
\begin{center}
\includegraphics[width=9cm]{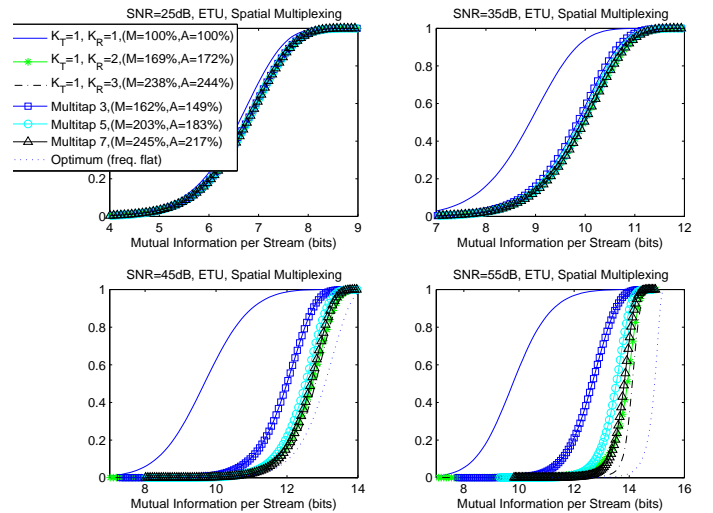}
\end{center}
\caption{Mutual information distribution (bits per stream) for different
levels of the background noise with $N_{T}=2$, $N_{R}=4$ and spatial
multiplexing (ETU\ channel model). }%
\label{fig:etu_2x4}%
\end{figure}

In this subsection, we evaluate the performance under background noise and
under a large set of randomly drawn channel frequency responses. The total
number of subcarriers was set to $1024$ and the number of antennas was fixed
to $N_{T}=2$ and $N_{R}=4$, and two different symbol streams were transmitted
$N_{S}=2$. The transmitter was fixed to $\mathbf{A}(\omega)=\mathbf{I}_{N_{T}%
}$ (pure spatial multiplexing) whereas a LMMSE\ processor was considered at
the receiver, i.e.
\[
\mathbf{B}(\omega)=\mathbf{H}(\omega)\left(  \mathbf{H}^{H}(\omega
)\mathbf{H}(\omega)+\sigma^{2}/P_{S}\mathbf{I}_{N_{T}}\right)  ^{-1}.
\]
Figs. \ref{fig:eva_2x4}\ and \ref{fig:etu_2x4} represent the cumulative
distribution function of the measured mutual information per stream
corresponding to $100$ \ realizations of EVA and ETU\ channel models
respectively, for different values of the signal to noise power ratio. These
mutual informations were estimated assuming Gaussian signaling and
disregarding the statistical dependence between distortion and information
symbols. Apart from the performance of the proposed receiver with multiple
parallel stages, we also represent the performance of the multi-tap MIMO
equalizer in \cite{Ihalainen2011}, based on the frequency sampling technique,
as well as the optimum performance under frequency flat equivalent channels.
In the legend of the figures, we represent the percentage of increase of the
corresponding technique in terms of real-valued multiplications (M\%)\ and
additions (A\%) with respect to the traditional single tap per-subcarrier
channel inversion (obtained as $K_{R}=1$). Observe that the parallel
multi-stage architecture with $K_{R}=2$ presents a computational complexity
that is comparable to a multi-tap processor with $N_{taps}=3$, but achieves a
much better output SNDR, especially at low values of the background noise. In
terms of the global SNDR distribution, two parallel stages are sufficient to
provide a performance comparable to a multi-tap filter with $N_{taps}=7$ taps
at a much lower computational complexity.

\begin{figure}[tbh]
\begin{center}
\includegraphics[width=9cm]{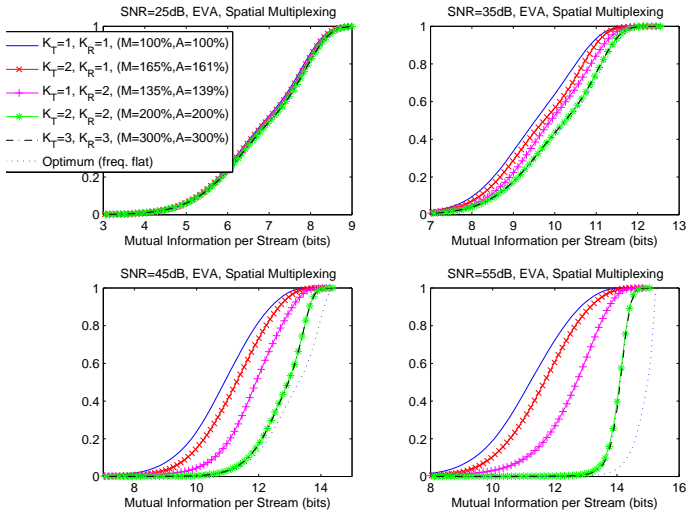}
\end{center}
\caption{Mutual information distribution (bits per stream) for different
levels of the background noise with $N_{T}=4$, $N_{R}=2$ and SVD-based
precoding (EVA\ channel model). }
\label{fig:eva_4x2}%
\end{figure}

\begin{figure}[tbhtbh]
\begin{center}
\includegraphics[width=9cm]{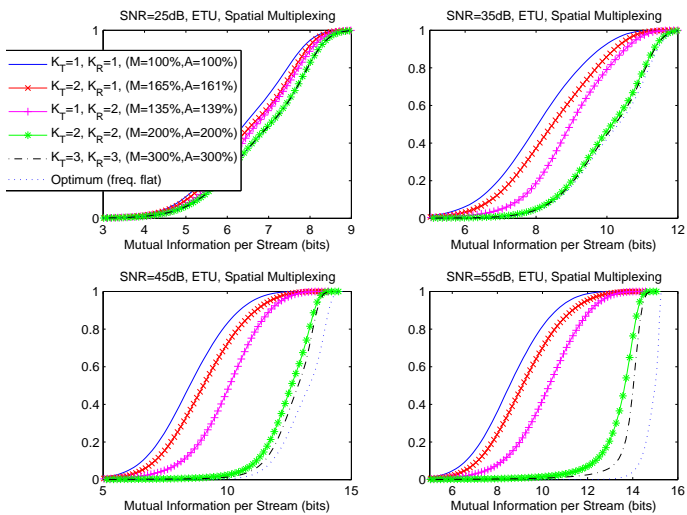}
\end{center}
\caption{Mutual information distribution (bits per stream) for different
levels of the background noise with $N_{T}=4$, $N_{R}=2$ and SVD-based
precoding (ETU\ channel model). }%
\label{fig:etu_4x2}%
\end{figure}

Next, we considered a scenario with $N_{T}=4$ and $N_{R}=2$ where the precoder
used the two left singular vectors associated with the largest singular values
of the channel matrix. The linear filter at the receiver was fixed in order to
invert the resulting channel matrix. Figs. \ref{fig:eva_4x2} to
\ref{fig:etu_4x2} show the distribution of the estimated mutual information
obtained with $100$ random realizations of the EVA and the ETU channel models
respectively and for different values of the background noise power. Here
again, we observe that high gains can be obtained with the proposed
multi-stage MIMO architecture using only two stages at the transmitter and at
the receiver.

\section{\label{sec:conclusions}Conclusions}

A novel parallel multi-stage MIMO architecture for FBMC transmissions under
strong frequency selectivity has been presented. The rationale behind the
approach consists in implementing a Taylor expansion of the ideal precoder and
linear receiver at the central frequency of each subband. By properly
exploiting the filterbank structure, it has been shown that the global system
can be implemented using conventional per-subcarrier precoders/linear
receivers in combination with parallel filterbanks constructed from sequential
derivatives of an original prototype pulse. For the specific case of
FBMC/OQAM, an asymptotic expression for the ICI/ISI distortion power has been
obtained. It has been shown that the global performance of the system
essentially depends on the minimum between the number of parallel stages
implemented at the transmitter and the receiver. Finally, numerical evaluation
studies indicate that the asymptotic performance assessment provides a very
accurate approximation of the reality for moderate values of the number of
subcarriers, and that significant gains can be obtained using the proposed
architecture under strong frequency selectivity.%

\appendices

\section{\label{sec:appendix_theorem_proof}Proof of Theorem \ref{th:Pe}}

We begin the proof by introducing a technical result that will be used
throughout this appendix, which will be separately proven.

\begin{proposition}
\label{prop:TaylorZ}Let $\mathcal{Y}_{p,q}\left(  \mathbf{S}_{n}\right)  $ be
the FBMC receive sample matrix corresponding to the symbol matrix
$\mathbf{S}_{n}$ when the channel is ideal and the prototype pulses $p,q$ are
used at the transmit/receive sides respectively. Let $F(\omega)$ denote a
$\mathcal{C}^{R^{\prime}+1}\left(  \mathbb{R}/2\pi\mathbb{Z}\right)  $
function for some integer $R^{\prime}>\left(  2R+1\right)  \left(  R+2\right)
$, and let $\mathcal{Z}_{p,q}^{F}\left(  \mathbf{S}_{n}\right)  $ denote the
matrix of received samples at the output of the decimators corresponding to
$\mathcal{Y}_{p,q}\left(  \mathbf{S}_{n}\right)  $, when the signal goes
through a channel with frequency response $F(\omega)$. Under $(\mathbf{As1}%
)-(\mathbf{As3})$ for any integer $R>0$ we can write%
\begin{align}
\mathcal{Z}_{p,q}^{F}\left(  \mathbf{S}_{n}\right)   &  =\sum_{r=0}^{R}%
\frac{\left(  -\textnormal{j}\right)  ^{r}}{r!\left(  2M\right)  ^{r}%
}\Lambda\left(  F^{\left(  r\right)  }\right)  \mathcal{Y}_{p,q^{(r)}}\left(
\mathbf{S}_{n}\right)  +o\left(  M^{-R}\right) \label{eq:prop_orders}\\
&  =\sum_{r=0}^{R}\frac{\left(  -\textnormal{j}\right)  ^{r}}{r!\left(
2M\right)  ^{r}}\mathcal{Y}_{p^{(r)},q}\left(  \Lambda\left(  F^{\left(
r\right)  }\right)  \mathbf{S}_{n}\right)  +o\left(  M^{-R}\right)
\label{eq:prop_orders2}%
\end{align}
where $F^{\left(  r\right)  }$ denotes the $r$th order derivative of the
function $F$ and where $o\left(  M^{-R}\right)  \,$\ for an integer $R$
denotes a matrix of potentially increasing dimensions whose entries decay to
zero faster than $M^{-R}\,$\ when $M\rightarrow\infty$. Furthermore, the above
identities hold true also if either $p$ in (\ref{eq:prop_orders}) or $q$ in
(\ref{eq:prop_orders2}) are replaced by $p^{(k)}$ and $q^{(k)}$ for any
integer $k\leq R$.
\end{proposition}

\begin{IEEEproof}
The identity in (\ref{eq:prop_orders}) is proven in \cite[Proposition
1]{mestre13tsp} when $F\left(  \omega\right)  $ is the Fourier transform of a
finite length sequence. The proof of the present result follows along the same
lines, see further Appendix \ref{sec:app_prop_proof}.
\end{IEEEproof}

\begin{corollary}
\label{corollary_FG}Under the above conditions, let $G(\omega)$ denote another
$\mathcal{C}^{R^{\prime}+1}\left(  \mathbb{R}/2\pi\mathbb{Z}\right)  $
function. Then,
\begin{align*}
\mathcal{Z}_{p,q}^{FG}\left(  \mathbf{S}_{n}\right)   &  =\sum_{k=0}^{R}%
\frac{\left(  -\textnormal{j}\right)  ^{k}}{k!\left(  2M\right)  ^{k}%
}\Lambda\left(  G^{\left(  k\right)  }\right)  \mathcal{Z}_{p,q^{(k)}}%
^{F}\left(  \mathbf{S}_{n}\right)  +o\left(  M^{-R}\right) \\
&  =\sum_{k=0}^{R}\frac{\left(  -\textnormal{j}\right)  ^{k}}{k!\left(
2M\right)  ^{k}}\mathcal{Z}_{p^{(k)},q}^{F}\left(  \Lambda\left(  G^{\left(
k\right)  }\right)  \mathbf{S}_{n}\right)  +o\left(  M^{-R}\right)  .
\end{align*}
Furthermore, the above identities also hold when the zeroth order derivatives
$p$ and $q$ are replaced by $p^{(k)}$ and $q^{(k)}$ for any integer $k\leq R$.
\end{corollary}

\begin{IEEEproof}
We will only prove the first identity, the proof of second one being
completely equivalent. Noting that $q^{(k)}\in\mathcal{C}^{R-k}$ and replacing
$\mathcal{Z}_{p,q^{(k)}}^{F}\left(  \mathbf{S}_{n}\right)  $ by the
corresponding expression in (\ref{eq:prop_orders}), with $R$ replaced by $R-k$
and $q$ replaced by $q^{(k)}$, we see that
\begin{align*}
&  \sum_{k=0}^{R}\frac{\left(  -\textnormal{j}\right)  ^{k}}{k!\left(
2M\right)  ^{k}}\Lambda\left(  G^{\left(  r\right)  }\right)  \mathcal{Z}%
_{p,q^{(k)}}^{F}\left(  \mathbf{S}_{n}\right) \\
&  =\sum_{k=0}^{R}\sum_{r=0}^{R-k}\frac{\left(  -\textnormal{j}\right)
^{k+r}}{k!r!\left(  2M\right)  ^{k+r}}\Lambda\left(  G^{\left(  r\right)
}F^{\left(  k\right)  }\right)  \mathcal{Y}_{p,q^{(k+r)}}\left(
\mathbf{S}_{n}\right) \\
&  +o\left(  M^{-R}\right) \\
&  \overset{\text{(a)}}{=}\sum_{m=0}^{R}\frac{\left(  -\textnormal{j}%
\right)  ^{m}}{m!\left(  2M\right)  ^{m}}\Lambda\left(  \left(  FG\right)
^{(m)}\right)  \mathcal{Y}_{p,q^{(m)}}\left(  \mathbf{S}_{n}\right)  +o\left(
M^{-R}\right) \\
&  \overset{\text{(b)}}{=}\mathcal{Z}_{p,q}^{FG}\left(  \mathbf{S}_{n}\right)
+o\left(  M^{-R}\right)
\end{align*}
where in (a) we have replaced the index $r$ by the index $m=k+r$ and swapped
the two sums and in (b) we have used again (\ref{eq:prop_orders}) with $F$
replaced by $FG$.\
\end{IEEEproof}

This corollary will be very useful in order to characterize the asymptotic
distortion error. Consider the expression of the received signal matrix
$\mathcal{Z}_{n}$ in (\ref{eq:receivedZ}), which can be expressed as%
\begin{align*}
\mathcal{Z}_{n}  &  =\sum_{n_{_{S}},n_{T},n_{R}}\sum_{\ell_{2}=0}^{K_{R}%
-1}\frac{\left(  -\textnormal{j}\right)  ^{\ell_{2}}}{\ell_{2}!\left(
2M\right)  ^{\ell_{2}}}\Lambda\left(  B_{n_{R},n}^{\left(  \ell_{2}\right)
\ast}\right)  \times\\
&  \times\sum_{\ell_{1}=0}^{K_{T}-1}\frac{\left(  -\textnormal{j}\right)
^{\ell_{1}}}{\ell_{1}!\left(  2M\right)  ^{\ell_{1}}}\mathcal{Z}_{p^{(\ell
_{1})},q^{(\ell_{2})}}^{H_{n_{R},n_{T}}}\left(  \Lambda\left(  A_{n_{T}%
,n_{_{S}}}^{\left(  \ell_{1}\right)  }\right)  \mathbf{S}_{n_{S}}\right)
\end{align*}
Applying Corollary \ref{corollary_FG} for $R=K_{T}+K_{R}$, we can write%
\begin{align*}
&  \sum_{\ell_{1}=0}^{K_{T}-1}\frac{\left(  -\textnormal{j}\right)
^{\ell_{1}}}{\ell_{1}!\left(  2M\right)  ^{\ell_{1}}}\mathcal{Z}_{p^{(\ell
_{1})},q^{(\ell_{2})}}^{H_{n_{R},n_{T}}}\left(  \Lambda\left(  A_{n_{T}%
,n_{_{S}}}^{\left(  \ell_{1}\right)  }\right)  \mathbf{S}_{n_{S}}\right) \\
&  =\mathcal{Z}_{p,q^{(\ell_{2})}}^{H_{n_{R},n_{T}}A_{n_{T},n_{_{S}}}}\left(
\mathbf{S}_{n}\right) \\
&  -\sum_{\ell_{1}=K_{T}}^{K_{T}+K_{R}}\frac{\left(  -\textnormal{j}%
\right)  ^{\ell_{1}}}{\ell_{1}!\left(  2M\right)  ^{\ell_{1}}}\mathcal{Z}%
_{p^{(\ell_{1})},q^{(\ell_{2})}}^{H_{n_{R},n_{T}}}\left(  \Lambda\left(
A_{n_{T},n_{_{S}}}^{\left(  \ell_{1}\right)  }\right)  \mathbf{S}_{n_{S}%
}\right) \\
&  +o\left(  M^{-\left(  K_{T}+K_{R}\right)  }\right)
\end{align*}
and therefore inserting this into the expression of $\mathcal{Z}_{n}$ above
and applying again Corollary \ref{corollary_FG} with respect to all the sums
in the index $\ell_{2}$, we obtain%
\[
\mathcal{Z}_{n}=\sum_{n_{_{S}},n_{R},n_{T}}\left[  \mathcal{Z}_{p,q}%
^{B_{n_{R},n}^{\ast}H_{n_{R},n_{T}}A_{n_{T},n_{_{S}}}}\left(  \mathbf{S}%
_{n_{S}}\right)  +\mathcal{E}_{n,n_{T},n_{R}}\left(  \mathbf{S}_{n_{S}%
}\right)  \right]
\]
where%
\begin{align*}
&  \mathcal{E}_{n,n_{T},n_{R}}\left(  \mathbf{S}_{n_{S}}\right) \\
&  =-\sum_{\ell_{2}=K_{R}}^{K_{T}+K_{R}}\frac{\left(  -\textnormal{j}%
\right)  ^{\ell_{2}}}{\ell_{2}!\left(  2M\right)  ^{\ell_{2}}}\Lambda\left(
B_{n_{R},n}^{\left(  \ell_{2}\right)  \ast}\right)  \mathcal{Z}_{p,q^{(\ell
_{2})}}^{H_{n_{R},n_{T}}A_{n_{T},n_{_{S}}}}\left(  \mathbf{S}_{n_{S}}\right)
\\
&  -\sum_{\ell_{1}=K_{T}}^{K_{T}+K_{R}}\frac{\left(  -\textnormal{j}%
\right)  ^{\ell_{1}}}{\ell_{1}!\left(  2M\right)  ^{\ell_{1}}}\mathcal{Z}%
_{p^{(\ell_{1})},q}^{B_{n_{R},n}^{\ast}H_{n_{R},n_{T}}}\left(  \Lambda\left(
A_{n_{T},n_{_{S}}}^{\left(  \ell_{1}\right)  }\right)  \mathbf{S}_{n_{S}%
}\right) \\
&  +\frac{\left(  -\textnormal{j}\right)  ^{K_{R}+K_{T}}}{K_{T}%
!K_{R}!\left(  2M\right)  ^{K_{R}+K_{T}}}\Lambda\left(  B_{n_{R},n}^{\left(
K_{R}\right)  \ast}\right)  \times\\
&  \times\mathcal{Z}_{p^{(K_{T})},q^{(K_{R})}}^{H_{n_{R},n_{T}}}\left(
\Lambda\left(  A_{n_{T},n_{_{S}}}^{\left(  K_{T}\right)  }\right)
\mathbf{S}_{n_{S}}\right)  +o\left(  M^{-\left(  K_{T}+K_{R}\right)  }\right)
.
\end{align*}
Now, using the linearity of the transmission from different antennas and the
fact that $\mathbf{B}^{H}(\omega)\mathbf{H}(\omega)\mathbf{A}(\omega
)=\mathbf{I}_{N_{s}}$ we obtain
\[
\mathcal{Z}_{n}=\mathcal{Y}_{p,q}\left(  \mathbf{S}_{n}\right)  +\sum
_{n_{T}=1}^{N_{T}}\sum_{n_{R}=1}^{N_{R}}\sum_{n_{_{S}}=1}^{N_{S}}%
\mathcal{E}_{n,n_{T},n_{R}}\left(  \mathbf{S}_{n_{S}}\right)
\]
Using now Proposition \ref{prop:TaylorZ} and disregarding all terms of higher
order, we can readily see that (\ref{eq_Sndecomposed})\ holds with%
\begin{equation}
\mathcal{E}_{p,q}\left(  \mathbf{S}_{n}\right)  =\mathcal{E}_{p,q}%
^{(1)}\left(  \mathbf{S}_{n}\right)  +\mathcal{E}_{p,q}^{(2)}\left(
\mathbf{S}_{n}\right)  +\mathcal{E}_{p,q}^{(3)}\left(  \mathbf{S}_{n}\right)
+o\left(  M^{-\left(  K_{T}+K_{R}\right)  }\right)  \label{eq:e_Sn}%
\end{equation}
where we have introduced the matrices
\begin{align}
\mathcal{E}_{p,q}^{(1)}\left(  \mathbf{S}_{n}\right)   &  =-\sum_{n_{_{S}}%
=1}^{N_{S}}\Upsilon_{n_{S}}\label{eq:error1}\\
\mathcal{E}_{p,q}^{(2)}\left(  \mathbf{S}_{n}\right)   &  =-\sum_{n_{_{S}}%
=1}^{N_{S}}\sum_{m=K_{R}}^{K_{T}+K_{R}}\Lambda\left(  \sum_{\ell_{2}=K_{R}%
}^{m}\beta_{n,n_{S}}^{(m,\ell)}\right)  \frac{\mathcal{Y}_{p,q^{(m)}}\left(
\mathbf{S}_{n_{_{S}}}\right)  }{\sqrt{2}\left(  2M\right)  ^{m}}%
\label{eq:error2}\\
\mathcal{E}_{p,q}^{(3)}\left(  \mathbf{S}_{n}\right)   &  =\sum_{n_{_{S}}%
=1}^{N_{S}}\Lambda\left(  \gamma_{n,n_{S}}\right)  \frac{\mathcal{Y}%
_{p^{(K_{T})},q^{(K_{R})}}\left(  \mathbf{S}_{n_{S}}\right)  }{\sqrt{2}\left(
2M\right)  ^{K_{R}+K_{T}}} \label{eq:error3}%
\end{align}
where $\beta_{n,n_{S}}^{(m,\ell)}$ and $\gamma_{n,n_{S}}$ are defined in
(\ref{eq:def_beta}) and (\ref{eq:def_gamma}) respectively, and where
\begin{multline*}
\Upsilon_{n_{S}}=\sum_{m=K_{T}}^{K_{T}+K_{R}}\frac{\left(  -\textnormal{j}%
\right)  ^{m}}{\left(  2M\right)  ^{m}m!}\sum_{n_{T}=1}^{N_{T}}\sum_{\ell
_{1}=K_{T}}^{m}\binom{m}{\ell_{1}}\times\\
\times\Lambda\left\{  \left(  \mathbf{B}^{H}\mathbf{H}\right)  ^{\left(
m-\ell_{1}\right)  }\right\}  _{n,n_{T}}\mathcal{Y}_{p^{(\ell_{1})}%
,q^{(m-\ell_{1})}}\left(  \Lambda\left(  A_{n_{T},n_{_{S}}}^{\left(  \ell
_{1}\right)  }\right)  \mathbf{S}_{n_{S}}\right)  .
\end{multline*}
Next, we transform $\Upsilon_{m,n_{S}}$ into a linear combination of matrices
of the type $\mathcal{Y}_{p^{(i)},q^{(j)}}\left(  \mathbf{S}_{n_{S}}\right)  $
for some integers $i,j$. The following lemma will be instrumental in this objective.

\begin{lemma}
\label{lemma_Conversion_order}Under the assumptions of Proposition
\ref{prop:TaylorZ}, we have
\begin{align*}
\mathcal{Y}_{p,q}\left(  \Lambda\left(  F\right)  \mathbf{S}_{n}\right)   &
=\sum_{m=0}^{R}\frac{\left(  -\textnormal{j}\right)  ^{m}}{\left(
2M\right)  ^{m}m!}\Lambda\left(  F^{\left(  m\right)  }\right)  \mathcal{Y}%
_{p,q}^{(m)}\left(  \mathbf{S}_{n}\right) \\
&  +o\left(  M^{-R}\right)
\end{align*}
where we have defined
\[
\mathcal{Y}_{p,q}^{(m)}\left(  \mathbf{S}_{n}\right)  =\sum_{r=0}^{m}\binom
{m}{r}\left(  -1\right)  ^{r}\mathcal{Y}_{p^{(r)},q^{(m-r)}}\left(
\mathbf{S}_{n}\right)  .
\]

\end{lemma}

\begin{IEEEproof}
For $\ell=0,\ldots,R$, we consider the identities in (\ref{eq:prop_orders}%
)-(\ref{eq:prop_orders2}) in Proposition \ref{prop:TaylorZ}\ with $p$ and $F$
replaced by $p^{(\ell)}$ and $F^{(\ell)}$ respectively, that is%
\begin{align*}
&  \sum_{r=0}^{R-\ell}\frac{\left(  -\textnormal{j}\right)  ^{r+\ell}%
}{r!\left(  2M\right)  ^{r+\ell}}\mathcal{Y}_{p^{(r+\ell)},q}\left(
\Lambda\left(  F^{\left(  r+\ell\right)  }\right)  \mathbf{S}_{n}\right) \\
&  =\sum_{r=0}^{R-\ell}\frac{\left(  -\textnormal{j}\right)  ^{r+\ell}%
}{r!\left(  2M\right)  ^{r+\ell}}\Lambda\left(  F^{\left(  r+\ell\right)
}\right)  \mathcal{Y}_{p^{(\ell)},q^{(r)}}\left(  \mathbf{S}_{n}\right)
+o\left(  M^{-R}\right)  .
\end{align*}
This forms a system of $R+1$ linear equations with $R+1$ unknowns, which can
be expressed in matrix form as
\[
\mathbb{A}_{R}\mathbf{x}_{R}=\mathbf{y}_{R}+o\left(  M^{-R}\right)
\]
where $\mathbf{x}_{R}=\left[  x_{0},\ldots,x_{R}\right]  ^{T}$, $\mathbf{y}%
_{R}=\left[  y_{0},\ldots,y_{R}\right]  ^{T}$,
\begin{align}
x_{r}  &  =\frac{\left(  -\textnormal{j}\right)  ^{r}}{\left(  2M\right)
^{r}}\mathcal{Y}_{p^{(r)},q}\left(  \Lambda\left(  F^{\left(  r\right)
}\right)  \mathbf{S}_{n}\right) \nonumber\\
y_{r}  &  =\sum_{m=0}^{R-r}\frac{\left(  -\textnormal{j}\right)  ^{m+r}%
}{m!\left(  2M\right)  ^{m+r}}\Lambda\left(  F^{\left(  m+r\right)  }\right)
\mathcal{Y}_{p^{(r)},q^{(m)}}\left(  \mathbf{S}_{n}\right)  \label{eq:yrdef}%
\end{align}
and where $\mathbb{A}_{R}$ is an $R\times R$ upper triangular Toeplitz matrix
with the entries of the $m$th upper diagonal fixed to $1/m$!, $m=0,\ldots,R$.
We are interested in obtaining the solution associated with the first entry of
$\mathbf{x}_{R}$, so that we will be able to write
\begin{equation}
x_{0}=\mathcal{Y}_{p,q}\left(  \Lambda\left(  F\right)  \mathbf{S}_{n}\right)
=\sum_{j=0}^{R}\xi_{j}y_{j}+o\left(  M^{-R}\right)
\label{eq:solution_toeplitz}%
\end{equation}
where $\xi_{j}$ are the entries of the upper row of $\mathbb{A}_{R}^{-1}$. We
can iteratively obtain the solution to $\xi_{j}$ by observing that we can
partition this matrix as
\[
\mathbb{A}_{R}=\left[
\begin{array}
[c]{cc}%
\mathbb{A}_{R-1} & \mathbf{J}_{R-1}\mathbf{a}_{R}\\
0 & 1
\end{array}
\right]
\]
where $\mathbf{a}_{R}=\left[  a_{1},\ldots,a_{R}\right]  ^{T}$, so that%
\[
\mathbb{A}_{R}^{-1}=\left[
\begin{array}
[c]{cc}%
\mathbb{A}_{R-1}^{-1} & -\mathbb{A}_{R-1}^{-1}\mathbf{J}_{R-1}\mathbf{a}_{R}\\
0 & 1
\end{array}
\right]
\]
and this basically implies that $\xi_{0}=1$ and
\[
\xi_{R}=-\sum_{m=0}^{R-1}\xi_{m}a_{R-m}=-\sum_{m=0}^{R-1}\frac{1}{(R-m)!}%
\xi_{m}.
\]
We can solve this recurrence by noting that it can be rewritten as
\[
R!\xi_{R}=-\sum_{m=0}^{R-1}\binom{R}{m}m!\xi_{m}%
\]
which basically implies that $m!\xi_{m}=\left(  -1\right)  ^{m}$. Using this
together with the expression of $y_{r}$ in (\ref{eq:yrdef}) and swapping the
two indexes we obtain the result of the lemma.%
\end{IEEEproof}

Applying Lemma \ref{lemma_Conversion_order} we can rewrite $\Upsilon_{n_{S}}$
as%
\begin{multline*}
\Upsilon_{n_{S}}=\sum_{m=K_{T}}^{K_{T}+K_{R}}\frac{\left(  -\textnormal{j}%
\right)  ^{m}}{m!\left(  2M\right)  ^{m}}\sum_{k=K_{T}}^{m}\sum_{\ell
_{1}=K_{T}}^{k}\binom{m}{k}\binom{k}{\ell_{1}}\times\\
\times\Lambda\left(  \left\{  \left(  \mathbf{B}^{H}\mathbf{H}\right)
^{\left(  k-\ell_{1}\right)  }\mathbf{A}^{\left(  \ell_{1}+m-k\right)
}\right\}  _{n,n_{S}}\right)  \mathcal{Y}_{p^{(\ell_{1})},q^{(k-\ell_{1})}%
}^{(m-k)}\left(  \mathbf{S}_{n_{S}}\right) \\
+o\left(  M^{-\left(  K_{T}+K_{R}\right)  }\right)
\end{multline*}
Using the fact that%
\begin{equation}
\binom{m}{k}\binom{k}{\ell_{1}}=\binom{m}{k-\ell_{1}}\binom{m-k+\ell_{1}}%
{\ell_{1}} \label{eq:identity_binomials}%
\end{equation}
and with the appropriate change of indexes ($\ell=m-k+\ell_{1}$), we see that
\begin{multline*}
\Upsilon_{n_{S}}=\sum_{m=K_{T}}^{K_{T}+K_{R}}\frac{1}{\sqrt{2}\left(
2M\right)  ^{m}}\Lambda\left(  \sum_{\ell=K_{T}}^{m}\alpha_{n,n_{S}}%
^{(m,\ell)}\right)  \times\\
\times\sum_{k=K_{T}+m-\ell}^{m}\binom{\ell}{m-k}\mathcal{Y}_{p^{(k-(m-\ell
))},q^{(m-\ell)}}^{(m-k)}\left(  \mathbf{S}_{n_{S}}\right)  +o\left(
M^{-\left(  K_{T}+K_{R}\right)  }\right)
\end{multline*}
where $\alpha_{n,n_{S}}^{(m,\ell)}$ is defined in (\ref{eq:def_alpha}).
Finally, using the change of indexes ($j=k-m+\ell+r$) together with
(\ref{eq:identity_binomials}), the additional change of indexes $s=\left(
j+m-\ell\right)  -k$ and the identity
\[
\sum_{s=0}^{j-K_{T}}\left(  -1\right)  ^{s}\binom{j}{s}=\left(  -1\right)
^{j-K_{T}}\binom{j-1}{j-K_{T}}%
\]
we finally obtain%
\begin{multline*}
\Upsilon_{n_{S}}=\sum_{m=K_{T}}^{K_{T}+K_{R}}\frac{1}{\sqrt{2}\left(
2M\right)  ^{m}}\Lambda\left(  \sum_{\ell=K_{T}}^{m}\alpha_{n,n_{S}}%
^{(m,\ell)}\right)  \times\\
\times\sum_{j=K_{T}}^{\ell}\left(  -1\right)  ^{j+K_{T}}\binom{\ell}{j}%
\binom{j-1}{K_{T}-1}\mathcal{Y}_{p^{(j)},q^{(m-j)}}\left(  \mathbf{S}_{n_{S}%
}\right) \\
+o\left(  M^{-\left(  K_{T}+K_{R}\right)  }\right)
\end{multline*}
Inserting this into the expression of $\mathcal{E}_{p,q}^{(1)}\left(
\mathbf{S}_{n}\right)  $ in (\ref{eq:error1}), we end up with an expression of
$\mathcal{E}_{p,q}\left(  \mathbf{S}_{n}\right)  $ in (\ref{eq:e_Sn}) that is
a linear combination of matrices of the form $\mathcal{Y}_{p^{(m)},q^{(n)}%
}\left(  \mathbf{S}_{n_{S}}\right)  $ for different integers $m,n$. Therefore,
we can analyze the asymptotic distortion variance by simply analyzing these
terms. We provide more details in what follows.

From the definition of the complex estimated symbols $\mathbf{\hat{s}}%
_{n}\left(  \ell\right)  $ in (\ref{eq:def_ahat}), we see that this column
vector is a function of two columns of the matrix $\mathcal{Z}_{p,q}\left(
\mathbf{S}_{n}\right)  $, namely
\begin{align*}
&  \mathbf{z}_{n,\ell}^{\text{odd}}\left(  p,q\right)  \overset
{\textnormal{def}}{=}\left[  \mathcal{Z}_{p,q}^{\text{odd}}\left(
\mathbf{S}_{n}\right)  \right]  _{:,\ell+\kappa-1}\text{ }\\
&  \mathbf{z}_{n,\ell}^{\text{even}}\left(  p,q\right)  \overset
{\textnormal{def}}{=}\left[  \mathcal{Z}_{p,q}^{\text{even}}\left(
\mathbf{S}_{n}\right)  \right]  _{:,\ell+\kappa}.
\end{align*}
Let us equivalently define $\mathbf{y}_{n,\ell}^{\text{odd}}\left(
p,q\right)  $ and $\mathbf{y}_{n,\ell}^{\text{even}}\left(  p,q\right)  $ as
above, replacing $\mathcal{Z}$ by $\mathcal{Y}$. We define the error
associated to the estimation of the $\ell$th multicarrier symbol of the $n$th
stream as $\mathbf{e}_{n,\ell}\left(  p,q\right)  =\mathbf{\hat{s}}_{n}\left(
\ell\right)  -\mathbf{s}_{n}\left(  \ell\right)  $, so that
\begin{equation}
\mathbf{e}_{n,\ell}\left(  p,q\right)  =\textnormal{Re}\left[
\mathbf{z}_{n,\ell}^{\text{odd}}\left(  p,q\right)  \right]
+\textnormal{j}\textnormal{Im}\left[  \mathbf{z}_{n,\ell}^{\text{even}%
}\left(  p,q\right)  \right]  -\mathbf{s}_{n}\left(  \ell\right) \nonumber
\end{equation}
Now, from the asymptotic description provided above we have been able to
express $\mathcal{Z}_{n}\left(  p,q\right)  $ as a function of\ matrices of
the form $\mathcal{Y}_{p^{(m)},q^{(k)}}\left(  \mathbf{S}_{n}\right)  $ when
$M\rightarrow\infty$ for several pairs of integers $m,k$. Consequently,
$\mathbf{\hat{s}}_{n}\left(  \ell\right)  $ is asymptotically described as a
weighted linear combination of $\mathbf{y}_{n,\ell}^{\text{odd}}\left(
p^{(m)},q^{(k)}\right)  $ and $\mathbf{y}_{n,\ell}^{\text{even}}\left(
p^{(m)},q^{(k)}\right)  \,$\ for several pairs of integers $m,k$. In order to
analyze the structure of $\mathcal{Y}_{n}$, let $\mathbf{F}$ denote the
$2M\times2M$ orthogonal Fourier matrix, and let $\mathbf{F}_{1}$ and
$\mathbf{F}_{2}$ be the matrices formed by selecting the $M$ upper and lower
rows of $\mathbf{F}$ respectively. The expression of $\mathcal{Y}%
_{n}^{\text{odd}}$ for FBMC/OQAM modulations can be shown to be
\cite{mestre13tsp}%
\begin{multline}
\mathcal{Y}_{p,q}^{\text{odd}}\left(  \mathbf{S}_{n}\right)  =2\mathbf{\Phi
F}^{H}\left(  \left[  \mathbf{F\Phi}^{\ast}\mathbf{B}_{n}\mathbf{,0,0}\right]
\circledast\mathcal{R}\left(  p,q\right)  \right) \label{eq_expr_Yeven}\\
+2\mathbf{\Phi F}^{H}\left(  \left[
\begin{array}
[c]{c}%
\mathbf{0},\mathbf{F}_{2}\mathbf{\Phi}^{\ast}\textnormal{j}\mathbf{C}%
_{n}\mathbf{,0}\\
\mathbf{F}_{1}\mathbf{\Phi}^{\ast}\textnormal{j}\mathbf{C}_{n}\mathbf{,0,0}%
\end{array}
\right]  \circledast\mathcal{S}\left(  p,q\right)  \right)
\end{multline}
where\ $\mathcal{R}\left(  p,q\right)  $, $\mathcal{S}\left(  p,q\right)  $
are defined in (\ref{def_R})-(\ref{def_S}), $\mathbf{\Phi}$ is a diagonal
matrix with its $m$th diagonal entry equal to $\exp\left(  -\textnormal{j}%
\pi\frac{M+1}{M}\left(  m-1\right)  \right)  $ and $\mathbf{0}$ is an
all-zeros column vector of appropriate dimensions. A similar expression can be
given for $\mathcal{Y}_{n}^{\text{even}}$, see further \cite[eq.
(4)]{mestre13tsp}.

Now, recalling that $\mathcal{I}$ is a $2M\times\left(  2\kappa-1\right)  $
matrix with ones in the central column and zeros elsewhere, we observe that we
are able to write
\begin{equation}
\mathbf{b}_{n}\left(  \ell\right)  =\left\{  2\mathbf{\Phi F}^{H}\left(
\left[  \mathbf{F\Phi}^{\ast}\mathbf{B}_{n}\mathbf{,0,0}\right]
\circledast\frac{1}{2}\mathcal{I}\right)  \right\}  _{:,\ell+\kappa-1}
\label{eq_identityBs}%
\end{equation}
and this identity holds true if we replace the pair $\mathbf{b}_{n}\left(
\ell\right)  $, $\mathbf{B}_{n}$ by $\mathbf{c}_{n}\left(  \ell\right)  $,
$\mathbf{C}_{n}$. Using (\ref{eq_identityBs}) and replacing $\mathbf{z}%
_{n,\ell}^{\text{odd}}\left(  p,q\right)  $ by the asymptotic expansion, we
see that%
\begin{align*}
&  \mathbf{z}_{n,\ell}^{\text{odd}}\left(  p,q\right)  -\mathbf{b}_{n}\left(
\ell\right) \\
&  =\mathbf{d}_{n,\ell}^{\text{odd}}\left(  p,q\right) \\
&  -\sum_{n_{_{S}}=1}^{N_{S}}\sum_{m=K_{T}}^{K_{T}+K_{R}}\frac{1}{\sqrt
{2}\left(  2M\right)  ^{m}}\sum_{\ell=K_{T}}^{m}\Lambda\left(  \alpha
_{n,n_{S}}^{(m,\ell)}\right)  \times\\
&  \times\sum_{j=K_{T}}^{\ell}\left(  -1\right)  ^{j-K_{T}}\binom{\ell}%
{j}\binom{j-1}{K_{T}-1}\mathbf{y}_{n_{S},\ell}^{\text{odd}}\left(
p^{(j)},q^{(m-j)}\right) \\
&  -\sum_{n_{_{S}}=1}^{N_{S}}\sum_{m=K_{R}}^{K_{T}+K_{R}}\frac{1}{\sqrt
{2}\left(  2M\right)  ^{m}}\Lambda\left(  \beta_{n,n_{S}}^{(m)}\right)
\mathbf{y}_{n_{S},\ell}^{\text{odd}}\left(  p,q^{(m)}\right) \\
&  +\frac{1}{\sqrt{2}\left(  2M\right)  ^{K_{R}+K_{T}}}\sum_{n_{_{S}}%
=1}^{N_{S}}\Lambda\left(  \gamma_{n,n_{S}}\right)  \mathbf{y}_{n_{S},\ell
}^{\text{odd}}\left(  p^{(K_{T})},q^{(K_{R})}\right) \\
&  +o\left(  M^{-\left(  K_{T}+K_{T}\right)  }\right)
\end{align*}
where $\mathbf{d}_{n,\ell}^{(\ast)}\left(  p,q\right)  $, $(\ast)\in\left\{
\text{odd,even}\right\}  $, is defined as $\mathbf{y}_{n,\ell}^{(\ast)}\left(
p,q\right)  $ by simply replacing $\mathcal{R}\left(  p,q\right)  $ with
$\mathcal{R}\left(  p,q\right)  -\frac{1}{2}\mathcal{I}$ in
(\ref{eq_expr_Yeven}). An equivalent expression can be derived for
$\mathbf{z}_{n,\ell}^{\text{even}}\left(  p,q\right)  -\textnormal{j}%
\mathbf{c}_{n}\left(  \ell\right)  $, which is omitted here due to space
constraints. The expressions presented in Theorem \ref{th:Pe} are obtained by
computing the variance of $\left\{  \mathbf{e}_{n,\ell}\left(  p,q\right)
\right\}  _{k}$ and disregarding the higher order terms. This can be easily
done using the following result, which can be proven as in \cite[Appendix
B]{mestre13tsp}.

\begin{lemma}
Consider now four generic prototype filters $p_{1},q_{1},p_{2},q_{2}$, and
denote $\mathcal{R}_{i}=\mathcal{R}\left(  p_{i},q_{i}\right)  $ and
$\mathcal{S}_{i}=\mathcal{S}\left(  p_{i},q_{i}\right)  $, $i=1,2$. Write for
compactness $\mathbf{\bar{y}}_{n,\ell,i}^{(\ast)}=\mathbf{\bar{y}}_{n,\ell
}^{(\ast)}\left(  p_{i},q_{i}\right)  \,$,$\ i=1,2$, $(\ast)\in\left\{
\text{even,odd}\right\}  $ and let $\ell\in\left\{  \kappa,\ldots
,N-\kappa\right\}  $. Under $\mathbf{(As4)}$, and for the FBMC/OQAM signal
model we can write%
\begin{align*}
\mathbb{E}\left[  \textnormal{Re}\left\{  \mathbf{y}_{n,\ell,1}^{\text{odd}%
}\right\}  _{k}\textnormal{Re}\left\{  \mathbf{y}_{n,\ell,2}^{\text{odd}%
}\right\}  _{k}\right]   &  =\mathbb{E}\left[  \textnormal{Im}\left\{
\mathbf{y}_{n,\ell,1}^{\text{even}}\right\}  _{k}\textnormal{Im}\left\{
\mathbf{y}_{n,\ell,2}^{\text{even}}\right\}  _{k}\right] \\
&  =\eta^{\left(  +,-\right)  }\left(  \mathcal{R}_{1}\mathcal{R}_{2}%
^{T},\mathcal{S}_{1}\mathcal{S}_{2}^{T}\right) \\
\mathbb{E}\left[  \textnormal{Im}\left\{  \mathbf{y}_{n,\ell,1}^{\text{odd}%
}\right\}  _{k}\textnormal{Im}\left\{  \mathbf{y}_{n,\ell,2}^{\text{odd}%
}\right\}  _{k}\right]   &  =\mathbb{E}\left[  \textnormal{Re}\left\{
\mathbf{y}_{n,\ell,1}^{\text{even}}\right\}  _{k}\textnormal{Re}\left\{
\mathbf{y}_{n,\ell,2}^{\text{even}}\right\}  _{k}\right] \\
&  =\eta^{\left(  -,+\right)  }\left(  \mathcal{R}_{1}\mathcal{R}_{2}%
^{T},\mathcal{S}_{1}\mathcal{S}_{2}^{T}\right) \\
\mathbb{E}\left[  \textnormal{Re}\left\{  \mathbf{y}_{n,\ell,1}^{\text{odd}%
}\right\}  _{k}\textnormal{Im}\left\{  \mathbf{y}_{n,\ell,2}^{\text{odd}%
}\right\}  _{k}\right]   &  =\mathbb{E}\left[  \textnormal{Re}\left\{
\mathbf{y}_{n,\ell,1}^{\text{even}}\right\}  _{k}\textnormal{Im}\left\{
\mathbf{y}_{n,\ell,2}^{\text{even}}\right\}  _{k}\right] \\
&  =0
\end{align*}
where we have defined for $s_{1},s_{2}\in\left\{  +,-\right\}  $,
\[
\eta^{\left(  s_{1},s_{2}\right)  }\left(  \mathcal{R}_{1}\mathcal{R}_{2}%
^{T},\mathcal{S}_{1}\mathcal{S}_{2}^{T}\right)  =\frac{P_{s}}{2M}%
\textnormal{tr}\left[  \mathcal{R}_{1}\mathcal{R}_{2}^{T}\mathbf{U}^{s_{1}%
}+\mathcal{S}_{1}\mathcal{S}_{2}^{T}\mathbf{U}^{s_{2}}\right]  .
\]
Furthermore, if $\mathbf{y}_{n,\ell,i}^{(\ast)}$ is replaced by $\mathbf{d}%
_{n,\ell,i}^{(\ast)}=\mathbf{d}_{n,\ell}^{(\ast)}\left(  p_{i},q_{i}\right)  $
in any of the above expressions, the same results hold replacing
$\mathcal{R}_{i}$ by $\mathcal{R}_{i}-\frac{1}{2}\mathcal{I}$.
\end{lemma}

\section{\label{sec:app_prop_proof}Proof of Proposition \ref{prop:TaylorZ}}

Let $\mathcal{Z}_{n}^{F}\left(  p,q\right)  $ denote the $2M\times2\left(
N+2\kappa\right)  $ matrix containing the received samples at the output of
the receive FFT corresponding to the $n$th transmit stream, assuming that the
transmit and receive prototype pulses are $p$ and $q$ respectively. For the
rest of the proof, we will drop the dependence on $n\,\ $in that matrix, and
we will decompose $\mathcal{Z}^{F}\left(  p,q\right)  =\mathcal{Z}%
_{\text{even}}^{F}\left(  p,q\right)  \otimes\left[  1,0\right]
+\mathcal{Z}_{\text{odd}}^{F}\left(  p,q\right)  \otimes\left[  0,1\right]
\,$. We will denote by $f\left[  \ell\right]  $ the $\ell$th coefficient of
the Fourier series of $F(\omega)$, i.e.%
\[
f\left[  \ell\right]  =\frac{1}{2\pi}\int_{0}^{2\pi}F\left(  \omega\right)
\textnormal{e}^{\textnormal{j}\omega\ell}d\omega.
\]
Furthermore, in order to describe the effect of the frequency selectivity of
$F\left(  \omega\right)  $, we introduce the following pulse-specific
matrices, defined for any $\ell\in\mathbb{Z}$ such that $-M<\ell\leq M$,%
\begin{align*}
\mathcal{R}_{\ell}\left(  p,q\right)   &  =\mathbf{P\circledast J}%
_{2M}\mathbf{Q}(\ell)=\left[
\begin{array}
[c]{c}%
\mathbf{P}_{1}\circledast\mathbf{J}_{M}\mathbf{Q}_{2}(\ell)\\
\mathbf{P}_{2}\circledast\mathbf{J}_{M}\mathbf{Q}_{1}(\ell)
\end{array}
\right] \\
\mathcal{S}_{\ell}\left(  p,q\right)   &  =\left[
\begin{array}
[c]{c}%
\mathbf{0},\mathbf{P}_{2}\circledast\mathbf{J}_{M}\mathbf{Q}_{2}(\ell)\\
\mathbf{P}_{1}\circledast\mathbf{J}_{M}\mathbf{Q}_{1}(\ell)\mathbf{,0}%
\end{array}
\right]  \text{ }%
\end{align*}
where $\mathbf{Q}(\ell)$ is defined as
\begin{align*}
\mathbf{Q}(\ell)  &  =\left[
\begin{array}
[c]{c}%
\mathbf{0},\mathbf{0},\left\{  \mathbf{Q}\right\}  _{2M-\ell+1:2M,:}\\
\mathbf{0},\left\{  \mathbf{Q}\right\}  _{1:2M-\ell,:},\mathbf{0}%
\end{array}
\right]  ,\text{ }0\leq\ell\leq M,\\
\mathbf{Q}(\ell)  &  =\left[
\begin{array}
[c]{c}%
\mathbf{0},\left\{  \mathbf{Q}\right\}  _{-\ell+1:2M,:},\mathbf{0}\\
\left\{  \mathbf{Q}\right\}  _{1:-\ell,:},\mathbf{0,0}%
\end{array}
\right]  ,\text{ }-M<\ell<0\text{ }%
\end{align*}
so that $\mathbf{Q}(0)=[\mathbf{0},\mathbf{Q]}$. Furthermore, given a column
vector of $2M$ entries $\mathbf{u}$, we define $\mathcal{M}(\mathbf{u})$ as
the $2M\times2M$ matrix
\[
\mathcal{M}(\mathbf{u})=\Phi\mathbf{F}^{H}\textnormal{diag}\left(
\mathbf{F}\Phi^{\ast}\mathbf{u}\right)
\]
where $\mathbf{F}$ is the $2M$ Fourier matrix, $\left\{  \mathbf{F}\right\}
_{ij}=\left(  2M\right)  ^{-1/2}\textnormal{e}^{\textnormal{j}%
2\pi\left(  i-1\right)  \left(  j-1\right)  /\left(  2M\right)  }$, $1\leq
i,j\leq2M$,\ and where $\Phi$ is a $2M\times2M$ diagonal matrix with entries
$\left\{  \Phi\right\}  _{kk}=\textnormal{e}^{-\textnormal{j}\pi\left(
M+1\right)  \left(  k-1\right)  /\left(  2M\right)  }$, $k=1,\ldots,2M$.

We will provide here the proof of (\ref{eq:prop_orders}), the proof of
(\ref{eq:prop_orders2}) following the same line of reasoning. Furthermore, we
will only show that (\ref{eq:prop_orders}) holds for the odd columns of
$\mathcal{Z}^{F}\left(  p,q\right)  $, namely $\mathcal{Z}_{\text{odd}}%
^{F}\left(  p,q\right)  $, since the proof for $\mathcal{Z}_{\text{even}}%
^{F}\left(  p,q\right)  $ is almost identical. Using the above definitions and
following \cite[eq. (14)-(15)]{mestre13tsp}, we can write\footnote{In the
following expression, matrices indexed by values that are either nonpositive
or higher than the matrix dimension should be understood as zero. Observe that
the number of terms of the sum in $\ell$ is, in fact, finite.}%
\begin{multline*}
\left\{  \mathcal{Z}_{\text{odd}}^{F}\left(  p,q\right)  \right\}  _{:,i}=\\
=2\sum_{\ell=-\infty}^{\infty}\sum_{j=1}^{N}f\left[  \ell\right]  \Theta
^{\ell}%
\Bigg[%
\mathcal{M}(\mathbf{b}_{j})\left\{  \mathcal{R}_{\left\langle \ell
\right\rangle _{2M}}\left(  p,q\right)  \right\}  _{:,i-j-\left[  \ell\right]
_{2M}+2}\\
+\mathcal{M}(\textnormal{j}\mathbf{c}_{j})\left\{  \mathcal{S}%
_{\left\langle \ell\right\rangle _{2M}}\left(  p,q\right)  \right\}
_{:,i-j-\left[  \ell\right]  _{2M}+2}%
\Bigg]
\end{multline*}
where $\Theta$ is a diagonal matrix with entries $\left\{  \Theta\right\}
_{kk}=\textnormal{e}^{-\textnormal{j}\omega_{k}}$, $k=1,\ldots,2M$,
$\left[  \ell\right]  _{2M}\ $returns the integer that is closest to
$\ell/(2M)$ (with the convention that \thinspace$\lbrack\left(  2m+1\right)
M]_{2M}=m$ when $m\in\mathbb{Z}$) and where $\left\langle \ell\right\rangle
_{2M}=\ell-[\ell]_{2M}$. Now, following the approach in \cite{mestre13tsp}, we
see that we can write%
\begin{align}
&  \left\{  \mathcal{Z}_{\text{odd}}^{F}\left(  p,q\right)  -\sum_{r=0}%
^{R}\frac{\left(  -\textnormal{j}\right)  ^{r}}{r!\left(  2M\right)  ^{r}%
}\Lambda\left(  F^{\left(  r\right)  }\right)  \mathcal{Y}_{\text{odd}}\left(
p,q^{(r)}\right)  \right\}  _{:,i}\nonumber\\
&  =2\sum_{\ell=-\infty}^{\infty}\sum_{j=1}^{N}f\left[  \ell\right]
\Theta^{\ell}%
\Bigg[%
\mathcal{M}(\mathbf{b}_{j})\left\{  \mathcal{E}_{\left\langle \ell
\right\rangle _{2M},R}^{1}\right\}  _{:,i-j-\left[  \ell\right]  _{2M}%
+2}\nonumber\\
&  +\mathcal{M}(\textnormal{j}\mathbf{c}_{j})\left\{  \mathcal{E}%
_{\left\langle \ell\right\rangle _{2M},R}^{2}\right\}  _{:,i-j-\left[
\ell\right]  _{2M}+2}%
\Bigg]
\label{eq:comberrorpulse}%
\end{align}
where we have defined, for $-M<\ell\leq M,$
\begin{align*}
\mathcal{E}_{\ell,R}^{1}  &  =\mathcal{R}_{\ell}\left(  p,q\right)
-\sum_{t=0}^{R}\frac{\left(  -\ell\right)  ^{t}}{t!\left(  2M\right)  ^{t}%
}\mathcal{R}_{0}\left(  p,q^{(t)}\right) \\
\mathcal{E}_{\ell,R}^{2}  &  =\mathcal{S}_{\ell}\left(  p,q\right)
-\sum_{t=0}^{R}\frac{\left(  -\ell\right)  ^{t}}{t!\left(  2M\right)  ^{t}%
}\mathcal{S}_{0}\left(  p,q^{(t)}\right)
\end{align*}
and where we have used the fact that (using the integration by parts formula)
\[
\frac{1}{2\pi}\int_{0}^{2\pi}F^{t}\left(  \omega\right)  \textnormal{e}^{\textnormal{j}\omega\ell}d\omega=\ell^{t}\left(
-\textnormal{j}\right)  ^{t}f\left[  \ell\right]  ,\ 0\leq t\leq R.
\]
Now, we separate the global sum in (\ref{eq:comberrorpulse}) into two terms,
which will be bounded in a different way. Consider a parameter $\delta
\in\left(  0,1\right)  $ and divide the sum with respect to $\ell$ in
(\ref{eq:comberrorpulse}) in two terms, corresponding to $\left\vert
\ell\right\vert <M^{\delta}$ and $\left\vert \ell\right\vert \geq M^{\delta}$.
Let us denote by $\chi_{1}$ and $\chi_{2}$ these two terms, so that
$(\ref{eq:comberrorpulse})=\chi_{1}+\chi_{2}$, where $\chi_{1}=\sum
_{\left\vert \ell\right\vert \geq M^{\delta}}($\textperiodcentered$)$ and
$\chi_{2}=\sum_{\left\vert \ell\right\vert <M^{\delta}}($\textperiodcentered
$)$. These two terms will be bounded using different methods, as it is
described next.

\subsection{Bounding the term $\chi_{1}$}

First observe that we can bound the $\left(  m,n\right)  $th entry of
$\mathcal{E}_{\ell,R}^{1}$ as
\begin{align*}
\left\vert \left\{  \mathcal{E}_{\ell,R}^{1}\right\}  _{m,n}\right\vert  &
\leq\left\vert \left\{  \mathcal{R}_{\ell}\left(  p,q\right)  \right\}
_{m,n}\right\vert \\
&  +\sum_{t=0}^{R}\frac{1}{t!}\left\vert \frac{\ell}{2M}\right\vert
^{t}\left\vert \left\{  \mathcal{R}_{0}\left(  p,q^{(t)}\right)  \right\}
_{m,n}\right\vert .
\end{align*}
Now, since the two pulses $p,q$ and their derivatives are bounded by
assumption, the absolute value of the entries of $\mathcal{R}_{\ell}\left(
p,q\right)  $ and $\mathcal{R}_{0}\left(  p,q^{(t)}\right)  $ are upper
bounded by a positive constant independent of $M$, denoted here by$\ C$.
Therefore, since $\left\vert \ell\right\vert <M$ in the definition of
$\mathcal{E}_{\ell,R}^{1}$,
\[
\left\vert \left\{  \mathcal{E}_{\ell,R}^{1}\right\}  _{m,n}\right\vert \leq
C+C\sum_{t=0}^{R}\frac{1}{t!}\left\vert \frac{\ell}{2M}\right\vert ^{t}\leq
C\left(  1+\sum_{t=0}^{R}\frac{2^{-t}}{t!}\right)
\]
which is bounded by a positive constant independent of $M$. A similar
reasoning can be applied to show that $\left\vert \left\{  \mathcal{E}%
_{\ell,R}^{2}\right\}  _{i,j}\right\vert $ has the same property. Therefore,
we see that (using the triangular and the Cauchy-Schwarz inequality),
\begin{multline*}
\left\vert \left\{  \chi_{1}\right\}  _{k,i}\right\vert \leq2\sum_{\left\vert
\ell\right\vert \geq M^{\delta}}\left\vert f\left[  \ell\right]  \right\vert
\\
\sum_{j=1}^{N}%
\Bigg[
\left\Vert \left\{  \mathcal{M}(\mathbf{b}_{j})\right\}  _{k,:}\right\Vert
\left\Vert \left[  \mathcal{E}_{\textnormal{smod}\left(  \ell,M\right)
,R}^{1}\right]  _{:,i-j+\left[  \ell/(2M)\right]  +2}\right\Vert +\\
+\left\Vert \left\{  \mathcal{M}(\textnormal{j}\mathbf{c}_{j})\right\}
_{k,:}\right\Vert \left\Vert \left[  \mathcal{E}_{\textnormal{smod}\left(
\ell,M\right)  ,R}^{2}\right]  _{:,i-j+\left[  \ell/(2M)\right]
+2}\right\Vert
\Bigg]
\\
\leq K_{1}\sqrt{M}\left(  \left\Vert \left\{  \mathcal{M}(\mathbf{b}%
_{j})\right\}  _{k,:}\right\Vert +\left\Vert \left\{  \mathcal{M}%
(\textnormal{j}\mathbf{c}_{j})\right\}  _{k,:}\right\Vert \right)
\sum_{\left\vert \ell\right\vert \geq M^{\delta}}\left\vert f\left[
\ell\right]  \right\vert \\
\leq K_{2}\sqrt{M}\sum_{\left\vert \ell\right\vert \geq M^{\delta}}\left\vert
f\left[  \ell\right]  \right\vert
\end{multline*}
for some positive constants $K_{1},K_{2}$ independent of $M$, where in the
last equation we have used the fact that $\left\Vert \left\{  \mathcal{M}%
(\mathbf{u})\right\}  _{k,:}\right\Vert $ is bounded above if the entries of
$\mathbf{u}$ are bounded (see further \cite[p.3604]{mestre13tsp}). Finally, we
need the following result:

\begin{lemma}
If $F\in\mathcal{C}^{R^{\prime}+1}\left(  \mathbb{R}/2\pi\mathbb{Z}\right)  $,
the $k$th Fourier coefficient $f[k]$ can be bounded by%
\[
\left\vert f[k]\right\vert \leq\frac{c}{\left\vert k\right\vert ^{R^{\prime
}+1}}%
\]
for some positive constant $c$.
\end{lemma}

\begin{IEEEproof}
Applying the partial integration formula to the definition $f\left[  k\right]
$ consecutively $R^{\prime}+1$ times,
\[
f\left[  k\right]  =\frac{\left(  -1\right)  ^{R^{\prime}+1}}{2\pi\left(
\textnormal{j}k\right)  ^{R^{\prime}+1}}\int_{0}^{2\pi}F^{\left(
R^{\prime}+1\right)  }\left(  \omega\right)  \textnormal{e}^{\textnormal{j}\omega k}d\omega
\]
and therefore the result follows by the triangular inequality for integrals,
taking $c=\sup_{\omega\in\mathbb{R}/2\pi\mathbb{Z}}\left\vert F^{R^{\prime}%
+1}\left(  \omega\right)  \right\vert $.%
\end{IEEEproof}

Applying this lemma, we readily see that there exists some positive constant
$K\,\ $such that, for any $\epsilon>0$,
\begin{multline*}
\left\vert \left\{  \chi_{1}\right\}  _{k,i}\right\vert \leq K\sqrt{M}%
\sum_{\ell\geq M^{\delta}}\frac{1}{\ell^{R^{\prime}+1}}\leq\\
\leq\frac{K}{M^{\delta\left(  R^{\prime}-\epsilon\right)  -1/2}}\sum_{\ell\geq
M^{\delta}}\frac{1}{\ell^{1+\epsilon}}=O\left(  M^{-\delta\left(  R^{\prime
}-\epsilon\right)  +1/2}\right)  .
\end{multline*}

\subsection{Bounding the term $\chi_{2}$}

In order to analyze this term, we will use the following result, which can be
proven as in \cite[Lemma 1]{mestre13tsp}.

\begin{lemma}
Let $\ell\in\mathbb{Z}$ be such that $\left\vert \ell\right\vert <M$. Then,
under $\mathbf{(As1)}$,
\[
\left\vert \left\{  \mathcal{E}_{\ell,R}^{1}\right\}  _{m,n}\right\vert \leq
K\left\vert \frac{\ell}{M}\right\vert ^{R+1}\text{, }\left\vert \left\{
\mathcal{E}_{\ell,R}^{2}\right\}  _{m,n}\right\vert \leq K\left\vert
\frac{\ell}{M}\right\vert ^{R+1}%
\]
for some positive constant $K$, independent of $M,m,n$ and $\ell$.
\end{lemma}

Using this, we readily see that, by the Cauchy-Schwarz inequality,
\begin{multline*}
\left\vert \left\{  \chi_{2}\right\}  _{k,i}\right\vert \leq2\sum_{\left\vert
\ell\right\vert <M^{\delta}}\left\vert f\left[  \ell\right]  \right\vert
\times\\
\sum_{j=1}^{N}\left\vert \left\{  \mathcal{M}(\mathbf{b}_{j})\right\}
_{k,:}\left[  \mathcal{E}_{\ell,R}^{1}\right]  _{:,i-j+2}\right\vert
+\left\vert \left\{  \mathcal{M}(\textnormal{j}\mathbf{c}_{j})\right\}
_{k,:}\left[  \mathcal{E}_{\ell,R}^{2}\right]  _{:,i-j+2}\right\vert \\
\leq K\sqrt{M}\sum_{\left\vert \ell\right\vert <M^{\delta}}\left\vert f\left[
\ell\right]  \right\vert \left\vert \frac{\ell}{M}\right\vert ^{R+1}=O\left(
M^{-\left(  1-\delta\right)  \left(  R+1\right)  +1/2}\right)
\end{multline*}
for some positive constant $K$.

\subsection{Concluding the proof}

With all the above, we have been able to show that the entries of
(\ref{eq:comberrorpulse}) are of the order $O(M^{-D})$, where
\[
D=\min\left\{  \left(  1-\delta\right)  \left(  R+1\right)  ,\delta\left(
R^{\prime}-\epsilon\right)  \right\}  -1/2
\]
for any $\delta\in\left(  0,1\right)  $ and $\epsilon>0$. As a function of
$\delta$, the maximum $D$ is obtained when
\[
\delta=\frac{\left(  R+1\right)  }{\left(  R^{\prime}-\epsilon+R+1\right)  }%
\]
and the corresponding exponent is given by
\[
D=\frac{\left(  R+1\right)  \left(  R^{\prime}-\epsilon\right)  }{\left(
R^{\prime}-\epsilon+R+1\right)  }-1/2.
\]
Now, if we require that $R^{\prime}>\left(  2R+1\right)  \left(  R+1\right)  $
and we fix $\epsilon\in\left(  0,R^{\prime}-\left(  2R+1\right)  \left(
R+1\right)  \right)  $, we have $D>R$, showing that $(\ref{eq:comberrorpulse}%
)=o\left(  M^{-R}\right)  $.

\bibliographystyle{ieeetr}

\end{document}